\documentclass[iop]{emulateapj}

\usepackage{hyperref}
\usepackage{graphicx}
\usepackage{natbib}

\newcommand{\xmm} {{\it XMM-Newton}}
\newcommand{\chandra} {{\it Chandra}}
\newcommand{\nustar} {{\it NuSTAR}}
\newcommand{\swift} {{\it Swift}}
\newcommand{\swiftxrt} {{\it Swift}/XRT}
\newcommand{\swiftbat} {{\it Swift}/BAT}

\newcommand{\cmsq} {cm$^{-2}$}
\newcommand{\nh} {$N_{\rm{H}}$}
\newcommand{\lx} {$L_{\rm{X}}$}

\newcommand{\chisq} {$\chi^2$}

\newcommand{\oiii}{{\rm{[O\,\sc{iii}]}}}
\newcommand{\hii}{{\rm{H\,\sc{ii}}}}
\newcommand{\nii}{{\rm{[N\,\sc{ii}]}}}

\newcommand{\ha}{{\rm{H$\alpha$}}}
\newcommand{\hb}{{\rm{H$\beta$}}}
\newcommand{\degree}{{$^\circ$}}
\newcommand{\ergs}{\mbox{\thinspace erg\thinspace s$^{-1}$}}
\newcommand{\ergcms}{\mbox{\thinspace erg\thinspace cm$^{-2}$\thinspace s$^{-1}$}}
\newcommand{\kms}{km\,s$^{-1}$}

\newcommand{\lbol} {$L_{\rm Bol}$}
\newcommand{\mbh} {$M_{\rm BH}$}

\newcommand{\lamedd} {$\lambda_{\rm Edd}$}

\newcommand{\msol} {$M_{\odot}$}

\shorttitle{ESO~509-IG066}
\shortauthors{Kosec et al.}

\begin{document}

\title{Investigating the evolution of the dual AGN system ESO~509-IG066}

\author{P. Kosec$^{1,2}$, M. Brightman$^{1}$, D. Stern$^{3}$, F. M\"{u}ller-S\'{a}nchez$^{4}$, M. Koss$^{5}$,  K. Oh$^{5}$, R. J. Assef$^{6}$, P. Gandhi$^{7}$, F. A. Harrison$^{1}$, H. Jun$^{3,1}$, A. Masini$^{8,9}$, C. Ricci$^{10,11,12}$, D. J. Walton$^{3,1}$, E. Treister$^{10}$, J. Comerford$^{4}$ and G. Privon$^{10}$}

\affil{$^{1}$Cahill Center for Astrophysics, California Institute of Technology, 1216 East California Boulevard, Pasadena, CA 91125, USA\\
$^{2}$Institute of Astronomy, Madingley Road, CB3 0HA Cambridge, UK\\
$^{3}$Jet Propulsion Laboratory, California Institute of Technology, Pasadena, CA 91109, USA\\
$^{4}$Department of Astrophysical and Planetary Sciences, University of Colorado, Boulder, CO 80309, USA\\
$^{5}$Institute for Astronomy, Department of Physics, ETH Zurich, Wolfgang-Pauli-Strasse 27, CH-8093 Zurich, Switzerland\\
$^{6}$N\'{u}cleo de Astronom\'{i}a de la Facultad de Ingenier\'{i}a, Universidad Diego Portales, Av. Ej\'{e}rcito Libertador 441, Santiago, Chile\\
$^{7}$Department of Physics and Astronomy, University of Southampton, Highfield, Southampton SO17 1BJ, UK\\
$^{8}$INAF Osservatorio Astronomico di Bologna, via Ranzani 1, I-40127 Bologna, Italy\\
$^{9}$Dipartimento di Fisica e Astronomia (DIFA), Universit\'{a} di Bologna, viale Berti Pichat 6/2, 40127 Bologna, Italy\\
$^{10}$Instituto de Astrof\'{i}sica, Facultad de F\'{i}sica, Pontificia Universidad Cat\'{o}lica de Chile, Casilla 306, Santiago 22, Chile\\
$^{11}$Kavli Institute for Astronomy and Astrophysics, Peking University, Beijing 100871, China\\
$^{12}$Chinese Academy of Sciences South America Center for Astronomy and China-Chile Joint Center for Astronomy, Camino El Observatorio 1515, Las Condes, Santiago, Chile\\}
\begin{abstract}

We analyze the evolution of the dual AGN in ESO~509-IG066, a galaxy pair located at $z=0.034$ whose nuclei are separated by 11~kpc. Previous observations with \xmm\ on this dual AGN found evidence for two moderately obscured (\nh$\sim10^{22}$ \cmsq) X-ray luminous (\lx$\sim10^{43}$ \ergs) nuclear sources. We present an analysis of subsequent \chandra, \nustar\ and \swiftxrt\ observations that show one source has dropped in flux by a factor of 10 between 2004 and 2011, which could be explained by either an increase in the absorbing column or an intrinsic fading of the central engine possibly due to a decrease in mass accretion. Both of these scenarios are predicted by galaxy merger simulations. The source which has dropped in flux is not detected by \nustar, which argues against absorption, unless it is extreme. However, new Keck/LRIS optical spectroscopy reveals a previously unreported broad \ha\ line which is highly unlikely to be visible under the extreme absorption scenario. We therefore conclude that the black hole in this nucleus has undergone a dramatic drop in accretion rate. From AO-assisted near-infrared integral-field spectroscopy of the other nucleus, we find evidence that the galaxy merger is having a direct effect on the kinematics of the gas close to the nucleus of the galaxy, providing a direct observational link between the galaxy merger and the mass accretion rate on to the black hole.

\end{abstract}


\section{Introduction}

Interactions between galaxies are predicted to cause increased nuclear activity \citep[e.g.][]{sanders88,hernquist89}. Massive gas flows triggered by gravitational interaction and resulting tidal forces can potentially fuel central supermassive black holes, creating luminous active galactic nuclei (AGN). This has been shown observationally in large statistical samples of galaxy pairs, where the AGN fraction and AGN luminosity have both been shown to increase as the separation between the galaxies decreases, peaking at $\sim10$ kpc \citep[][]{alonso07,woods07,ellison11,silverman11,koss12,satyapal14}. In addition, it is naturally expected that such a gas build-up in the nucleus will not only fuel the growth of the super-massive black hole, but obscure it as well, at Compton-thick levels \citep[\nh$>10^{24}$ \cmsq,][]{hopkins05}. Indeed, this has been shown recently with a sample of interacting galaxies at $z\sim1$ where galaxies that exhibit evidence for a merger or interaction were more likely to host a Compton-thick AGN than a less obscured one \citep[][]{kocevski15,ricci17}. 

However, while both observations and simulations of galaxy mergers find that AGN obscuration increases in galaxy mergers, simulations have also shown that large fluctuations in mass accretion rate on to the black hole are also to be expected \citep[e.g.][]{dimatteo05,hopkins06}, especially during the later stages of the merger \citep[e.g.][]{gabor15}. Observationally, however, it is challenging to distinguish between changes in accretion rate and changes in absorption since both lead to changes in the observed flux \citep[e.g.][]{rivers15b,gandhi17}.

\begin{figure}
\begin{center}
\includegraphics[width=90mm]{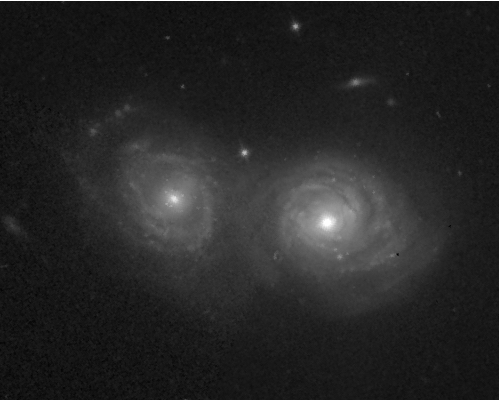}

\caption{{\it HST}/WFPC2 F606W image of the galaxy pair ESO~509-IG066 from \cite{malkan98}. The Eastern galaxy is on the left and the Western galaxy is on the right. The image is 50\arcsec$\times$40\arcsec. }
\label{hst_image}
\end{center}
\end{figure}

In this paper we study a local pair of interacting galaxies, ESO~509-IG066 (Figure \ref{hst_image}) located at $z=0.034$ ($D_{\rm L}\sim$150 Mpc). The galaxy pair was chosen for this study as part of a \nustar\ program to observe \swiftbat\ detected AGN \citep{harrison13}. The nuclei of the two galaxies are separated by 16\arcsec\ on the sky, which at this redshift implies a physical projected separation of 10.9~kpc \citep[assuming $H_{0}=67.8$ km~s$^{-1}$~Mpc$^{-1}$, $\Omega_{\rm m}=0.308$ and $\Omega_{\Lambda}=0.692$,][]{planck15}. The galaxies are aligned in the East-West direction where the RA and Dec of the nuclei are 203.6653\degree, $-$23.4468\degree, henceforth known as the ``Western source'' and 203.6700\degree, $-$23.4461\degree, henceforth known as the ``Eastern source''. \cite{guainazzi05} (G05) analyzed the galaxy pair using \xmm\ data from 2004 and reported that both galaxies host luminous nuclear X-ray sources with luminosities of $\sim10^{43}$ \ergs. They found that the Western source is a moderately obscured AGN with a column density of $\sim10^{23}$~\cmsq, while the Eastern source is almost unobscured with the column density less than 10$^{22}$ \cmsq. While the Western source is very weak in the very soft X-ray band (0.5$-$2 keV), it outshines the Eastern source in harder bands (2.0$-$10.0 keV). The system has also been detected by {\it Swift}/BAT \citep[][]{cusumano10,baumgartner13} with a 14$-$195 keV flux of 1.4$\times10^{-11}$ \ergcms, although with a PSF of 10\arcmin, \swiftbat\ cannot resolve these two nuclei. Furthermore, ESO~509-IG066 was detected by MAXI/GSC \citep{hiroi11} with a 4$-$10 keV flux of 1.7$\times10^{-11}$ \ergcms.

In this paper we reanalyze the \xmm\ data from 2004 and add new results from the \chandra, \nustar\ and \swift\ observations. In addition to the X-ray data, we use data from the Catalina Sky Survey \citep{drake09} and the {\it Wide-field Infrared Survey Explorer} \citep[{\it WISE},][]{wright10} to compare the X-ray variations with the variability in the optical and infrared (IR). Furthermore, we present new Keck/LRIS optical spectroscopic observations of the galaxies and a Keck/OSIRIS near-IR integral field spectroscopic observation of the Western nucleus that yield insights into the system. 

The structure of this paper is as follows. In Section \ref{sec_data} we describe the observational data used and the data reduction, Section \ref{sec_analysis} briefly summarizes our X-ray spectral fitting methods and results, followed by results from optical and IR analysis listed in Section \ref{sec_multiwav}. We present new Keck/LRIS optical spectroscopy and Keck/OSIRIS near-IR integral field spectroscopy in Section \ref{sec_spectro}. We discuss our results in Section \ref{sec_disc} and conclude in Section \ref{sec_conc}.

\section{Observations}
\label{sec_data}

\begin{figure*}
\begin{center}
\includegraphics[width=180mm]{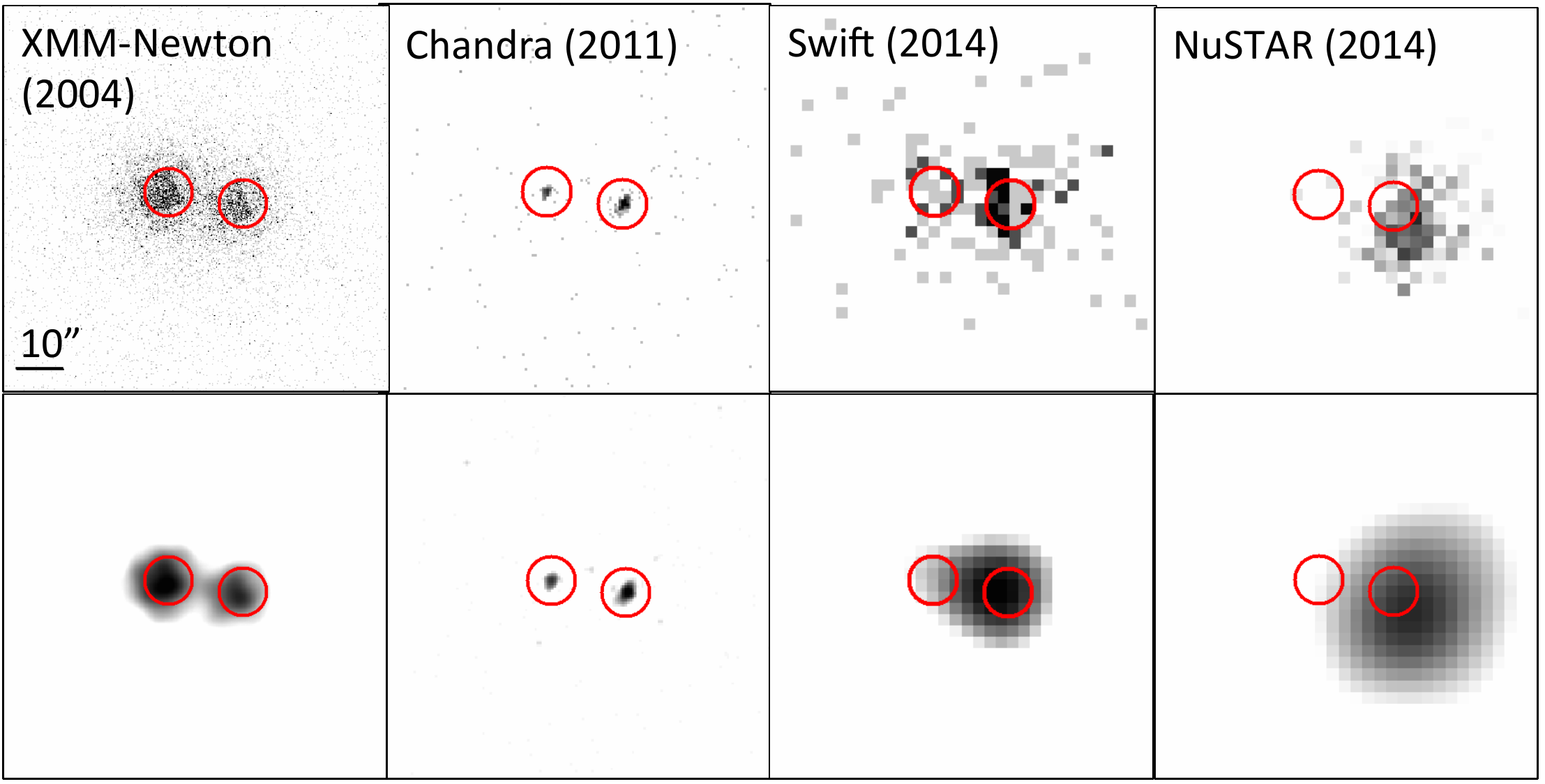}
\caption{\xmm\ (0.2-10 keV), \chandra\ (0.5-8 keV), \swiftxrt\  (0.5-10 keV) and \nustar\ (3-79 keV) images of ESO~509-IG066 from 2004-2014 showing the progressive fading of the Eastern source (East is left in these images). The red circles mark the positions of the sources and have 5\arcsec\ radii. All images have the same scale which is marked on the \xmm\ image. The top panels show the unsmoothed images and the bottom images show images that have been smoothed with a Gaussian kernel with radius 5\arcsec\ for \xmm, \swiftxrt\ and \nustar\ and 2\arcsec\ for \chandra.}
\label{Xray_image}
\end{center}
\end{figure*}

ESO~509-IG066 has been observed by \xmm, \chandra, \nustar\ and \swift, where the \nustar\ and \swift\ observations were simultaneous. Figure \ref{Xray_image} presents the X-ray images of the system from each of the observatories and Table \ref{table_obsdat} summarizes the basic observational data. The following sections discuss the processing of each of these X-ray data sets, as well as ancillary data sets at optical and IR wavelengths.

\begin{table}
\centering
\caption{X-ray observations.}
\label{table_obsdat}
\begin{center}
\begin{tabular}{l r r r r r}
\hline
Telescope & ObsID & Date & Exposure (ks)\\
(1) & (2) & (3) & (4) \\
\hline
\xmm&0200430801&2004-01-24&13.9\\
\chandra&12835&2011-02-08&5.1\\
\nustar&60061244002&2014-09-02&20.9\\
\swiftxrt&00080115002&2014-09-03&6.2\\
\hline
\end{tabular}
\tablecomments{Column (1) gives the telescope name, column (2) lists the observation ID, column (3) gives the start date of the observation and column (4) gives the exposure time in ks. }
\end{center}
\end{table}

\subsection{\xmm}

\xmm\ \citep{jansen01} EPIC-pn \citep{struder01} data were reduced using {\sc sas} v14.0, selecting events from a circular region of radius 60\arcsec\ centered on the galaxy pair corresponding to a $\sim90$\% encircled energy fraction. EPIC-MOS data were not considered due to their lower hard X-ray sensitivity. A period of high background at the beginning of the observation was filtered out, leaving 7.9 ks of science data. Background spectra were extracted from a nearby circular region of 75\arcsec\ radius on the same chip as the galaxies. Initially, both of the nuclei were extracted in a single spectrum. During subsequent analysis, we also extracted a spectrum for each of the two objects. For this individual analysis we used circular regions of radius 8\arcsec\ for the Eastern source and 7\arcsec\ for the Western source respectively. Spectra were grouped with a minimum of 20 counts per bin. We carried out spectral fitting in the 0.2$-$10.0 keV energy range.
%pattern 0 events
\subsection{\chandra}

\chandra\ \citep{weisskopf99} data of the Eastern and Western sources were extracted using the {\sc ciao} (v4.7, CALDB v4.6.5) tool {\sc specextract}, from circular regions with a radius of 5\arcsec. A larger circular region on the same chip as the galaxies was used to extract the background spectrum. The spectrum of the Western source (the brighter one) was grouped to at least 10 counts per bin and the spectrum of the Eastern source was binned to at least 5 counts per bin. Counts at energies below 0.5 and above 7.5 keV were ignored as the efficiency of the instrument drops quickly when out of this energy range.

\subsection{\nustar}

The \nustar\ \citep{harrison13} raw data were reduced using the {\sc nustardas} v 1.5.1 software. Initially the events were cleaned and filtered with the {\tt nupipeline} script with standard parameters, then the {\tt nuproducts} procedure was used to extract spectra and the corresponding response and auxiliary files. A single spectrum was extracted for the galaxy pair because the size of PSF of \nustar\ \citep[$\sim60$\arcsec,][]{madsen15} is larger than the separation of the galaxies. The spectra were extracted from circular regions centered on the peak of emission and with specific radii to maximize the signal-to-noise ratio. The background spectra were obtained from regions chosen to cover as much area as possible on the same detector as the source while avoiding the source itself and its point-spread function. Data from both focal plane modules (FPMA and FPMB) were extracted and used in simultaneous fitting without coadding. Both \nustar\ spectra were grouped by at least 20 counts per bin using the {\sc heasarc} tool {\sc grppha}. We ignore channels below 3 keV as the calibration at lower energies is uncertain, and channels above the 79 keV cut off that results from absorption in the mirror coating.

\subsection{\swift}

The {\it Swift}/XRT \citep{gehrels04,burrows05} observation was taken simultaneously with the {\it NuSTAR} observation. The data were preprocessed and the spectrum extracted using automatic routines {\sc xrtpipeline} and {\sc xrtproducts} before downloading. Because of the low spatial resolution of XRT (HPD=18\arcsec\ at 1.5 keV), only one spectrum was extracted for the AGN pair. We used default parameters (such as extraction radius) while generating the spectrum. The data were then grouped by at least 3 counts per bin. We carry out spectral fitting in the 0.2$-$7.0 keV energy range where the efficiency of the telescope is highest.

\subsection{Keck}

We obtained observations of the ESO~509-IG066 system with the Keck telescope during 2016. Optical spectroscopy of both nuclei was carried out using the Keck~I telescope and the dual-beam Low Resolution Imaging Spectrometer \citep[LRIS,][]{oke95}. The 300~s spectrum, obtained on UT 2016 June 9 in photometric conditions, used the 1\farcs5 wide slit, the 5600~\AA\ dichroic to split the light, the 600 $\ell$ mm$^{-1}$ grism on the blue arm ($\lambda_{\rm blaze} = 4000$~\AA), and the 600 $\ell$ mm$^{-1}$ grating on the red arm ($\lambda_{\rm blaze} = 7500$~\AA). The 1\farcs5 slit corresponds to physical scales of $\sim$1 kpc. The observations were obtained at a position angle of 82\degree\ in order to simultaneously observe both galaxies in the system.  We processed the data using standard techniques within IRAF, and calibrated the spectrum using standard stars observed using the same instrument configuration on the same night.

In addition to the optical spectroscopy, we acquired near-IR integral field spectroscopy of the nucleus of the Western galaxy from the adaptive optics (AO)-assisted near-IR integral-field spectrograph \citep[OSIRIS,][]{larkin06, vandam06, wizinowich06} on the Keck~I Telescope taken on UT 2016 April 22. The data were taken in the $K$-band using the $Kbb$ filter and the 0\farcs1 pixel scale, resulting in a rectangular field-of-view of 1\farcs6$\times$6\farcs4. The galaxy nucleus was used as tip-tilt star for the Laser Guide Star AO system.  A total of two sky and four on-source exposures of 600 s each at a position angle of 90 degrees were combined to make the final data cube. 

The OSIRIS data were reduced using the OSIRIS data reduction pipeline (ODRP). This performs all the usual steps needed to reduce near-IR spectra, but with the additional routines for reconstructing the data cube. More details can be found in \cite{msanchez16}. Flux calibration was performed using an A7V star HD 87035 ($K=7.5$).

\subsection{\it Other data}

We used {\it WISE} \citep{wright10} and {\it NEOWISE-R} \citep{mainzer11} data to investigate the IR variability of the galaxy pair, which are spatially resolved by the telescope. ESO~509-IG066 was observed three times by {\it WISE} in 2010$-$2011 and four times by NEOWISE in 2014$-$2016. Data from the Catalina Sky Survey  were used to investigate the variability of the AGN in the optical part of spectrum.

\section{X-ray spectral fitting}
\label{sec_analysis}

We fit the X-ray data using {\sc xspec} \citep{arnaud96} software version 12.9.0 and used the Cash \citep{cash79} statistic for fitting because of the low number of counts per bin in the \swift\ data. Both AGN are modeled in the same way with an absorbed cut-off power-law plus {\tt pexrav} \citep{magdziarz95} component simulating scattered radiation from the dusty torus plus a narrow iron line at 6.4 keV. We take into account Galactic absorption with a {\tt wabs} model component, the \nh\ value of which was obtained from the Leiden/Argentine/Bonn survey of Galactic HI \citep{kalberla05}, and found to be $6.67\times10^{20}$ \cmsq. We also included cross-normalization constants between the X-ray instruments. A secondary power-law component was added, assuming that a fraction of the primary radiation escapes through a patchy absorber without reprocessing, or is scattered into the line of sight. In {\sc xspec}, this model is written: {\tt constant*wabs*(constant*cutoffpl + zwabs*cabs(cutoffpl+pexrav+zgauss))}. The {\tt zwabs*cabs} component represents the reprocessing of the X-rays by photo-electric absorption and Compton scattering local to the source. The {\tt constant*cutoffpl} component represents the secondary power-law component. The {\tt pexrav} component represents scattered radiation from the torus. All the statistical errors calculated by {\sc XSPEC} are at 90 percent confidence level, unless explicitly stated otherwise.

\begin{figure}
\begin{center}
\includegraphics[width=90mm]{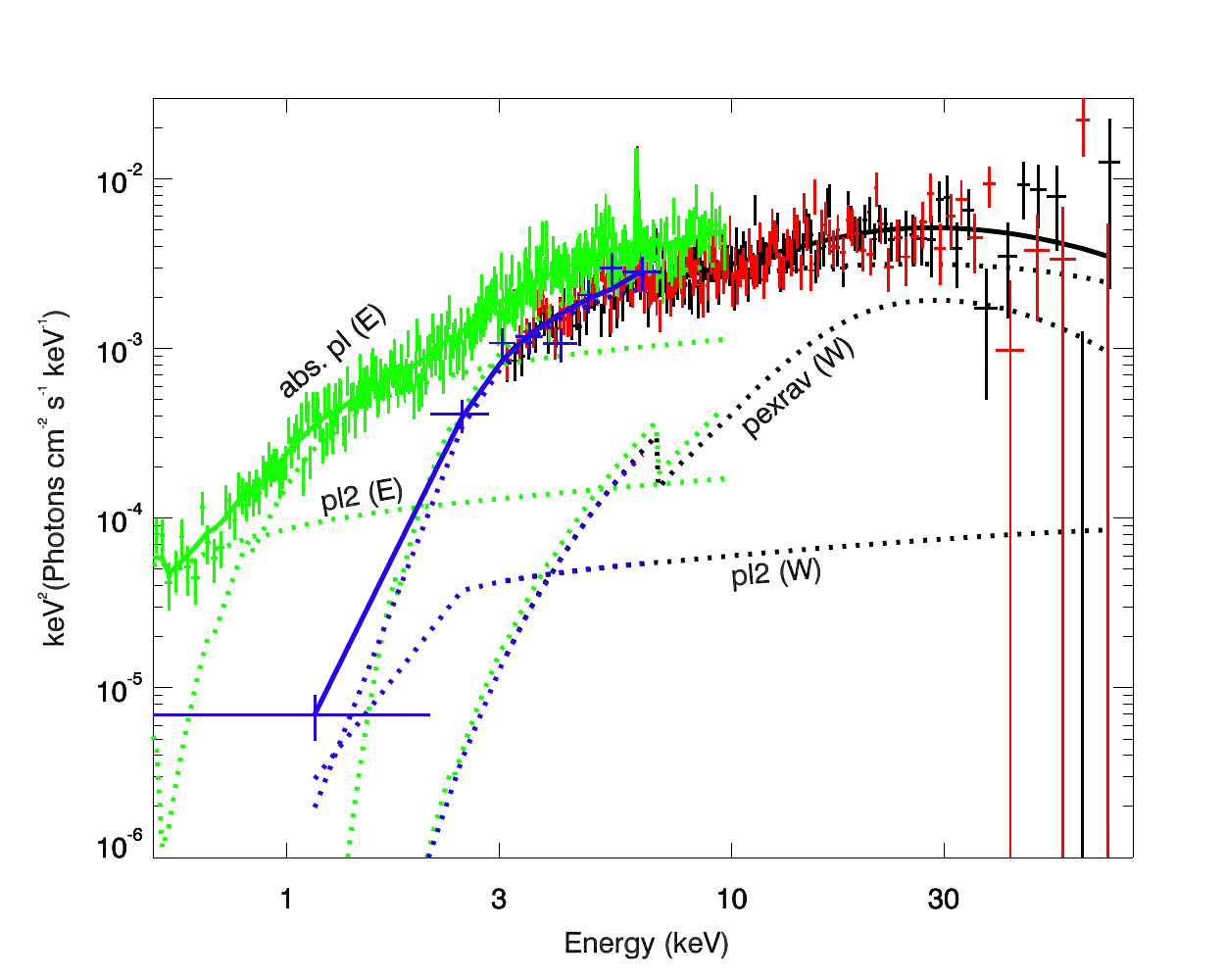}
\caption{$EF_{\rm E}$ spectra of the two sources from \xmm\ (green), \nustar\ (black and red) and \swiftxrt\ (blue), where both sources have been included in the extraction regions and the spectra have been unfolded through the instrumental responses. The absorbed power-law (abs. pl), secondary power-law (pl2) and pexrav components used in the fit are marked for each epoch.}
\label{fig_spectrum_total}
\end{center}
\end{figure}

Initially, we extracted a single spectrum for both of the sources from the \xmm, \swift\ and \nustar\ data and fitted the three data sets simultaneously. Since G05 showed that the two sources have different spectral properties, specifically the level of absorption, we assumed that we could spectrally decompose the two nuclei in the summed spectra. To do so we used the model described above multiplied by two in order to account for both AGN within the extraction region. At first we tied the parameters for each source across data sets under the assumption that they did not change between the \xmm\ observation and the \swift\ plus \nustar\ observations, however, the resulting fit was very poor. A visual inspection of the spectrum revealed that the 2004 \xmm\ spectrum was significantly different from the 2014 \nustar\ and \swiftxrt\ spectra, appearing harder (see Figure \ref{fig_spectrum_total}). For this reason, another constant component was applied to the components of one of the sources to account for possible variability of one source with respect to the other. From this we obtained a very good fit with a C-stat of 556.21 from 589 degrees of freedom (DOF). We found that the constant for one of the sources drops to 0 for the \nustar\  and \swiftxrt\ data implying that it is negligible in the 2014 data. The results of this fit are summarized in Table \ref{table_m2a}. We identify the source with the highest \nh\ value as the Western source and the other the Eastern source, since these match the parameters from G05 who carried out spatially resolved analysis on the galaxies.

\begin{table*}
\centering
\caption{Results of simultaneous fitting of both sources using \nustar, \swift\ and \xmm\ data.}
\label{table_m2a}
\begin{center}
\begin{tabular}{l r r r r r r r}
\hline
Source 	& \nh\  				& $\Gamma$ 			& $E_{\rm C}$ 	& Power-law norm.  					& $f_{\rm pl2}$ 				&Pexrav norm. 					& Iron line norm. \\
(1) & (2) & (3) & (4) & (5) & (6) & (7) & (8)\\
\hline
\\
West		& 7.3$_{-1.4}^{+1.5}$	& 1.65$_{-0.20}^{+0.23}$	& 71$^{+150}_{-30}$	& 1.56$_{-0.52}^{+0.68}\times10^{-3}$	& 6$_{-l}^{+25}\times10^{-3}$	& 1.5$_{-1.2}^{+2.8}\times10^{-3}$	& 1.21$_{-0.38}^{+0.42}\times10^{-5}$	\\
\\
East		& 0.60$_{-0.15}^{+0.22}$	& 1.84$_{-0.32}^{+0.62}$	& 500 		& 5.9$_{-1.2}^{+1.5}\times10^{-4}$		& 0.11$_{-l}^{+1.65}$	&-----------&------------\\
\\
\hline
\end{tabular}
\tablecomments{Column (1) gives source name, column (2) gives the \nh\ value in units of $10^{22}$~\cmsq, column (3) gives the photon index of the cut-off power-law, column (4) shows the exponential cutoff energy of the cut-off power-law in keV. Here `$-l$' signifies that the lower limit on this parameter is unconstrained, in this case consistent with zero. Column (5) lists the normalization of the power-law in units of photons keV$^{-1}$ cm$^{-2}$ s$^{-1}$ at 1 keV, column (6) gives the fraction of the leaked power-law model to the primary one ('-l' indicates that this fraction is unconstrained at the lower end), column (7) shows the normalization of the {\tt pexrav} component in units of  photons keV$^{-1}$ cm$^{-2}$ s$^{-1}$ at 1 keV, and column (8) lists the normalization of Gaussian component representing the iron line at 6.4 keV in units of total photons cm$^{-2}$ s$^{-1}$ in the line. The equivalent width of this line is 150 eV.}
\end{center}
\end{table*}

The first constant in the spectral model is used to account for differences between instruments and possible variability effects. We fix it to 1 for \nustar\ FPMA and let it float for all other instruments. Constants obtained by fitting the spectrum simultaneously are: 1.01$\pm$0.04 for FPMB, 1.26$_{-0.31}^{+0.22}$ for \xmm-pn and 1.01$_{-0.13}^{+0.15}$ for \swiftxrt. The cross-normalization for \xmm\ is higher, possibly due to variability between 2004 and 2014. However it is still consistent with unity within the 90 percent uncertainties. Noticing this, we investigated variability of the Western source. We tried freeing the photon index and column densities of this source in \xmm\ data, though the resulting values stayed constant within the uncertainties. The other cross-normalization constants are consistent with calibrated values from \cite{madsen15}.

\begin{figure}
\begin{center}
\includegraphics[width=90mm]{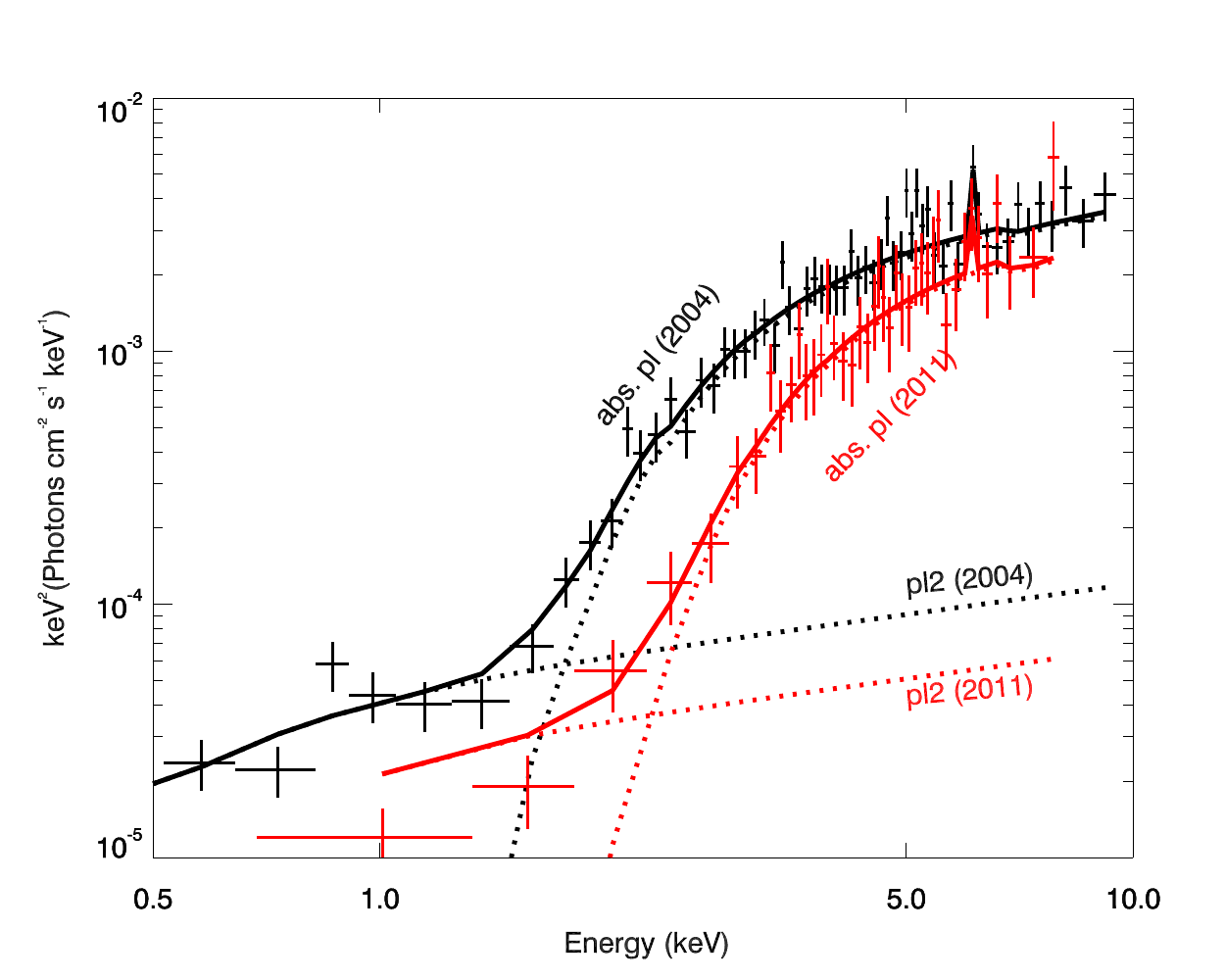}
\caption{$EF_{\rm E}$ spectra of the Western source from \xmm\ (black) and \chandra\ (red) where the spectra have been unfolded through the instrumental responses. The absorbed power-law (abs. pl) and secondary power-law (pl2) components used in the fit are marked for each epoch.}
\label{fig_SourceW}
\end{center}
\end{figure}

To investigate the variability of the two sources individually, we analyzed the \chandra\ observation from 2011. For \xmm, while the 16\arcsec\ separation of the sources is similar to the FWHM of the telescope's PSF, we extracted the spectra of each source using small 7$-$8\arcsec\ radius apertures following G05, bearing in mind that each spectrum will be contaminated by the other's PSF wings. First we extracted \xmm\ and \chandra\ spectra of the Western source only. We simultaneously fit the spectra using a simple spectral model in the form {\tt constant*wabs(constant*po + zwabs*cabs(po + zgauss))}. The power-law did not require a cut-off and the {\tt pexrav} component was not required since these features are not significant in the soft X-ray range (below 10 keV). As the \chandra\ observation is only 5 ks, we were not able to constrain all the parameters of the spectral model. We achieved a very good fit of 69.49/92 C-stat/DOF when freeing $f_{\rm pl2}$ (the fraction of the secondary power-law to the primary one), \nh\ and the cross-normalization constant, and fixing all other parameters. The results of the fit are shown in Table \ref{table_specpar} and in Figure \ref{fig_SourceW}. We notice an increase in the \nh\ of the absorber from 6$\pm1\times10^{22}$~\cmsq\ to 1.2$\pm0.2\times10^{23}$~\cmsq, though the uncertainties increased as well. The equivalent width of the Fe K$\alpha$ line is 82$^{+95}_{-78}$ eV during the \xmm\ observation and 68$^{+162}_{-68}$ eV during the \chandra\ observation. The cross-normalization constant between the measurements is consistent with unity at the 90 percent confidence level.

\begin{figure}
\begin{center}
\includegraphics[width=90mm]{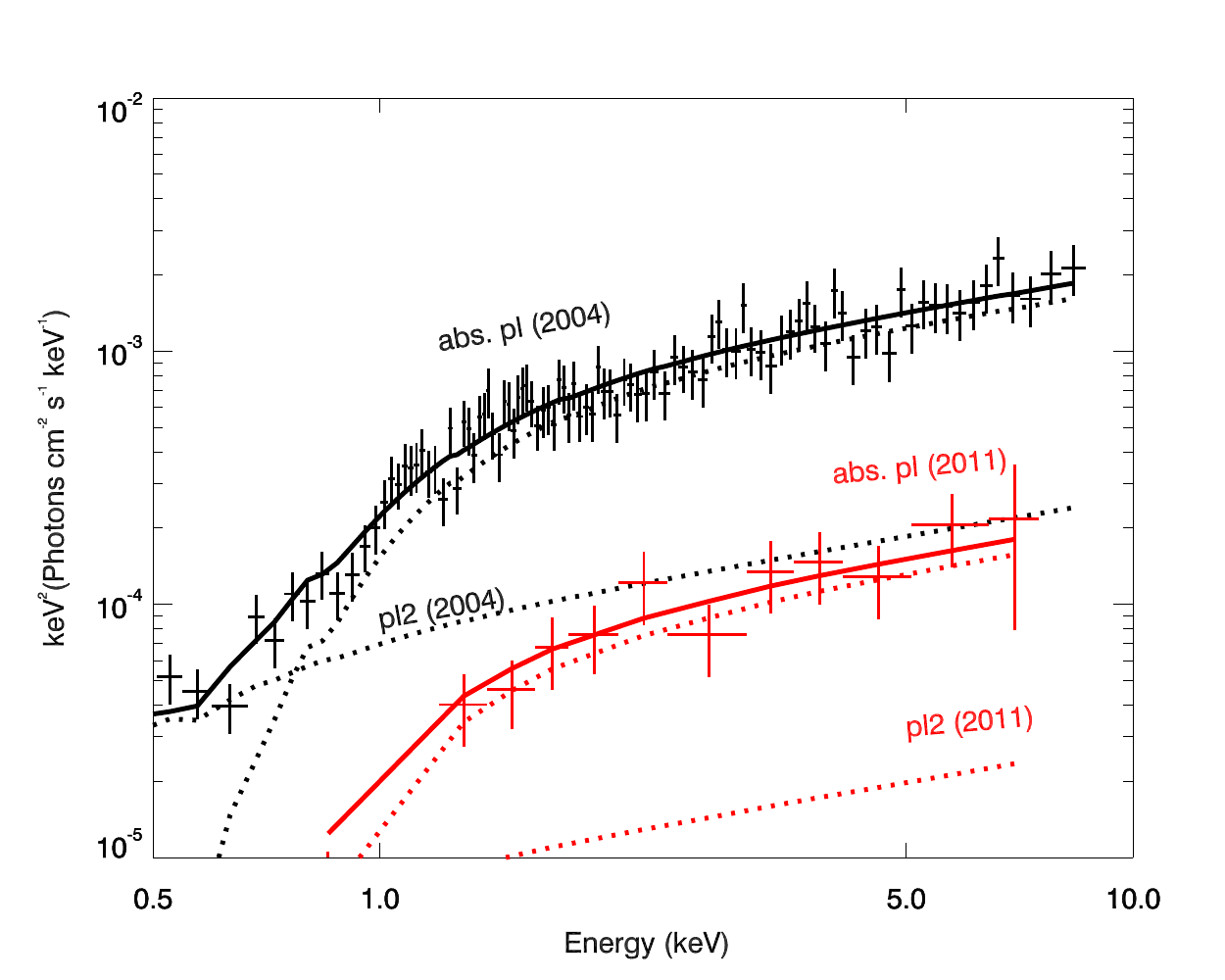}
\caption{$EF_{\rm E}$ spectra of the Eastern source from \xmm\ (black) and \chandra\ (red) where the spectra have been unfolded through the instrumental responses. The absorbed power-law (abs. pl) and secondary power-law (pl2) components used in the fit are marked for each epoch.}
\label{fig_SourceE}
\end{center}
\end{figure}

We extracted the spectrum of the Eastern source from \xmm\ and \chandra\ observations as done with the Western source. The source is detected in both observations, but the flux from the 2011 \chandra\ observation is much lower than from the 2004 \xmm\ observation. We fit the spectrum simultaneously using the same model as for the Western source. In this case, the {\tt zgauss} component is not significant anymore and again the {\tt pexrav} component and high-energy cut-off were not required. We obtained a  good fit with 85.18/115 C-stat/DOF by tying all the parameters for \chandra\  to the \xmm\ parameters except for the cross-normalization constant, which now accounts mostly for the drop in flux of the object. Upon freeing the other \chandra\ spectral parameters, the fit becomes unconstrained. The results are summarized in Table \ref{table_specpar} and Figure \ref{fig_SourceE}. We find that the normalization of the Eastern source decreased by a factor of $\sim$10 between 2004, when \xmm\ observed it and 2011, the date of the \chandra\ observation.

\begin{table*}
\centering
\caption{Simultaneous fitting results of both sources using \xmm\ and \chandra\ data.}
\label{table_specpar}
\begin{center}
\begin{tabular}{l c c c c c c }
\hline
Source 	& \nh\  				& $\Gamma$ 			& Power-law norm.  					& $f_{\rm pl2}$ 				& Iron line norm. 				& Cross-normalization constant \\
(1) & (2) & (3) & (4) & (5) & (6) & (7) \\
\hline
\multicolumn{7}{c}{Western source}\\
\xmm	&6.41$_{-0.80}^{+0.84}$	&1.53$\pm0.20$	&1.49$_{-0.46}^{+0.66}\times10^{-3}$	&0.031$_{-0.010}^{+0.015}$	&8.3$_{-7.1}^{+8.1}\times10^{-6}$	&fixed to 1\\
\chandra	&12.4$_{-1.6}^{+1.7}$	&tied					&tied								&0.011$_{-0.005}^{+0.007}$	&tied							&0.88$_{-0.14}^{+0.17}$\\
\multicolumn{7}{c}{Eastern source}\\
\xmm	&0.53$_{-0.12}^{+0.13}$	&1.49$\pm0.12$	&5.57$_{-0.92}^{+1.09}\times10^{-4}$	& 0.148$_{-0.042}^{+0.056}$	& -----------						&fixed to 1\\
\chandra	&tied					&tied					&tied								&tied						&-----------						&0.107$_{-0.017}^{+0.019}$\\
\hline
\end{tabular}
\tablecomments{Column (1) lists the instrument that measured the data, column (2) gives the \nh\ value in units of 10$^{22}$~\cmsq, column (3) gives the photon index of the power-law, column (4) lists the normalization of the power-law in units of photons keV$^{-1}$ cm$^{-2}$ s$^{-1}$ at 1 keV, column (5) gives the fraction of the secondary power-law to the primary one, column (6) gives the normalization of the {\tt zgauss} component in units of total photons cm$^{-2}$ s$^{-1}$ in the line, and column (7) shows the cross-normalization constant between the two measurements.}
\end{center}
\end{table*}

In the final part of X-ray analysis, we focus on fluxes from both of the sources and their change over time. We calculate the flux using  {\tt cflux}, which provides the observed flux of the combined spectral models. We calculate the flux in two different bands: soft (0.5$-$2.0 keV) and hard (2.0$-$8.0 keV). From the \chandra\ and \xmm\ data we obtain fluxes of the sources separately, while for \swift\ plus \nustar\ data, the calculated flux is the sum of both sources. However, we list it as the flux of Western source only since the Eastern source is undetected. We include the upper limit at 90 percent confidence level for the Eastern source flux calculated using spectral modeling. We plot the results from this analysis in Figure \ref{fig_lightcurves}.

The system was reported by {\it Swift}/BAT as having 14$-$195 keV flux of 1.4$\times10^{-11}$ \ergcms\ \citep[][]{cusumano10,baumgartner13}, which is an average over the period 2005$-$2009. We estimate the 14$-$195 keV flux of ESO 509-IG066 from the 2014 \nustar\ observation by extrapolating our spectral model up to 195 keV. We find that the flux during the \nustar\ observation is 1.5$\times10^{-11}$ \ergcms, which is consistent with the Swift/BAT flux reported, implying there is no evidence for a drop in X-ray flux in this band. However, this is not surprising since the Western source, which has remained relatively constant, dominates at high energies.

\section{Long-term optical and IR light curves}
\label{sec_multiwav}

To investigate the cause of the X-ray variability of the Eastern source, we examined the optical and IR lightcurves. We used the Catalina Real-Time Transient Survey (CRTS) to determine the $V$-band optical brightness of the Eastern source over time. The measurements were taken from August 2005 until July 2013 so the survey covers a large time period between our X-ray observations. 

We then analyzed the mid-IR brightness of both sources using {\it WISE} and {\it NEOWISE} data as described in Section \ref{sec_data}. The sources were observed during seven epochs, three times in 2010$-$2011 and four times in 2014$-$2016. The reduced \chisq\ of the standard profile fitting technique for {\it WISE} point sources listed in the AllWISE catalog indicates that the sources are extended. For this reason, we use small-aperture photometry, extracted from regions of 5.5\arcsec\ also listed in the AllWISE catalog.  Lastly, since the W3 and W4 passbands are not available for {\it NEOWISE}, we do not include them in our analysis. We present this photometry in Table \ref{table_WISE}.

\begin{table*}
\centering
\caption{WISE and NEOWISE photometry of both galaxies.}
\label{table_WISE}
\begin{center}
\begin{tabular}{l c c c c}
\hline
				&\multicolumn{2}{c}{Eastern source}			&\multicolumn{2}{c}{Western source}	\\
				& W1			& W2				& W1				& W2\\
(1) & (2) & (3) & (4) & (5) \\				
\hline
AllWISE 1 & 12.67$\pm$0.02 & 12.39$\pm$0.06 &11.79$\pm$0.01 &11.10$\pm$0.04 \\
AllWISE 2 & 12.75$\pm$0.02 &12.52$\pm$0.01& 11.83$\pm$0.02 &11.19$\pm$0.04 \\
AllWISE 3 & 12.89$\pm$0.04 &12.68$\pm$0.06 &11.86$\pm$0.03 &11.25$\pm$0.12 \\
NEOWISE-R 1 & 12.71$\pm$0.02 &12.53$\pm$0.03 &11.75$\pm$0.08 &11.07$\pm$0.03 \\
NEOWISE-R 2 & 12.83$\pm$0.03 &12.62$\pm$0.05 &11.73$\pm$0.02 &11.04$\pm$0.04 \\
NEOWISE-R 3 & 12.78$\pm$0.02 &12.59$\pm$0.03 &11.71$\pm$0.03 &11.05$\pm$0.08 \\ 
NEOWISE-R 4 & 12.70$\pm$0.03 &12.40$\pm$0.05 &11.72$\pm$0.05 &11.03$\pm$0.08 \\

\hline
\end{tabular}
\tablecomments{Column (1) shows the observational epoch and  columns (2)-(5) list the {\it WISE} and {\it NEOWISE} magnitudes (Vega) of both galaxies in the W1 and W2 bands.}
\end{center}
\end{table*}

Firstly we note that the mid-IR colors of the Eastern galaxy are relatively blue, with W1-W2$\simeq0.3$, which indicates that the bands are dominated by stellar emission. For the Western galaxy W1-W2$\simeq0.7$, which is more consistent with being dominated by the AGN \citep{stern12}. Therefore any drop in mid-IR flux from the AGN in the Eastern galaxy will probably be washed out by the host galaxy. While the flux in W1 from the Eastern galaxy does shows a $\sim20$\% drop during the 2010$-$2011 {\it WISE} observations and the AllWISE catalog gives it the maximum probability that the flux was not constant with time, the drop in flux is not sustained, as seen in the $2014-2016$ {\it NEOWISE} data which show a recovery of the initial flux. Finally, the expected optical to K-band time lag using our assumed cosmology, is 25 days, a much shorter timescale than the cadence of the {\it WISE} data (although the W1, W2 emission regions may be slightly larger than the K-band emission region).

\begin{figure}
\begin{center}
\includegraphics[width=100mm]{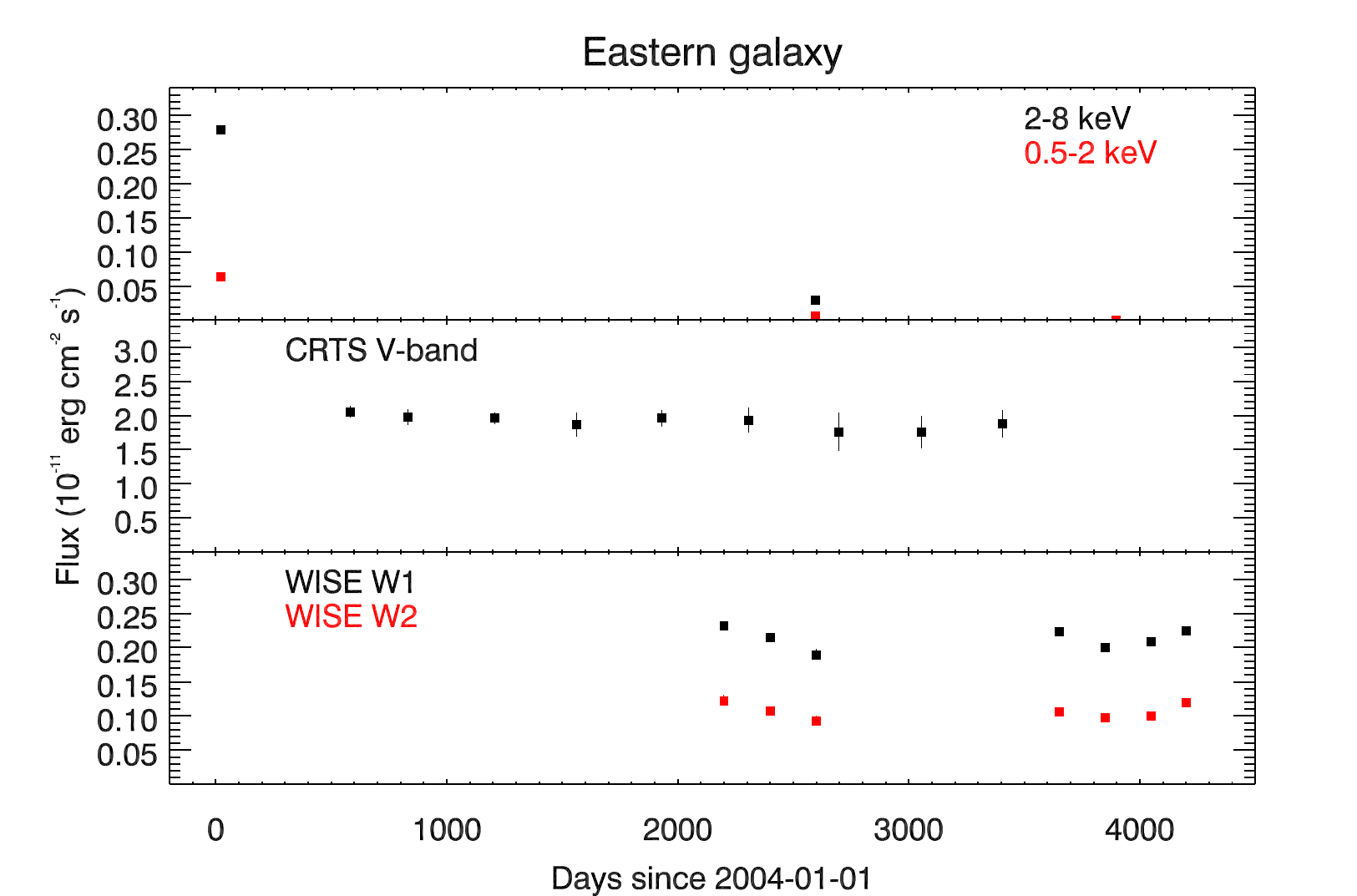}
\includegraphics[width=100mm]{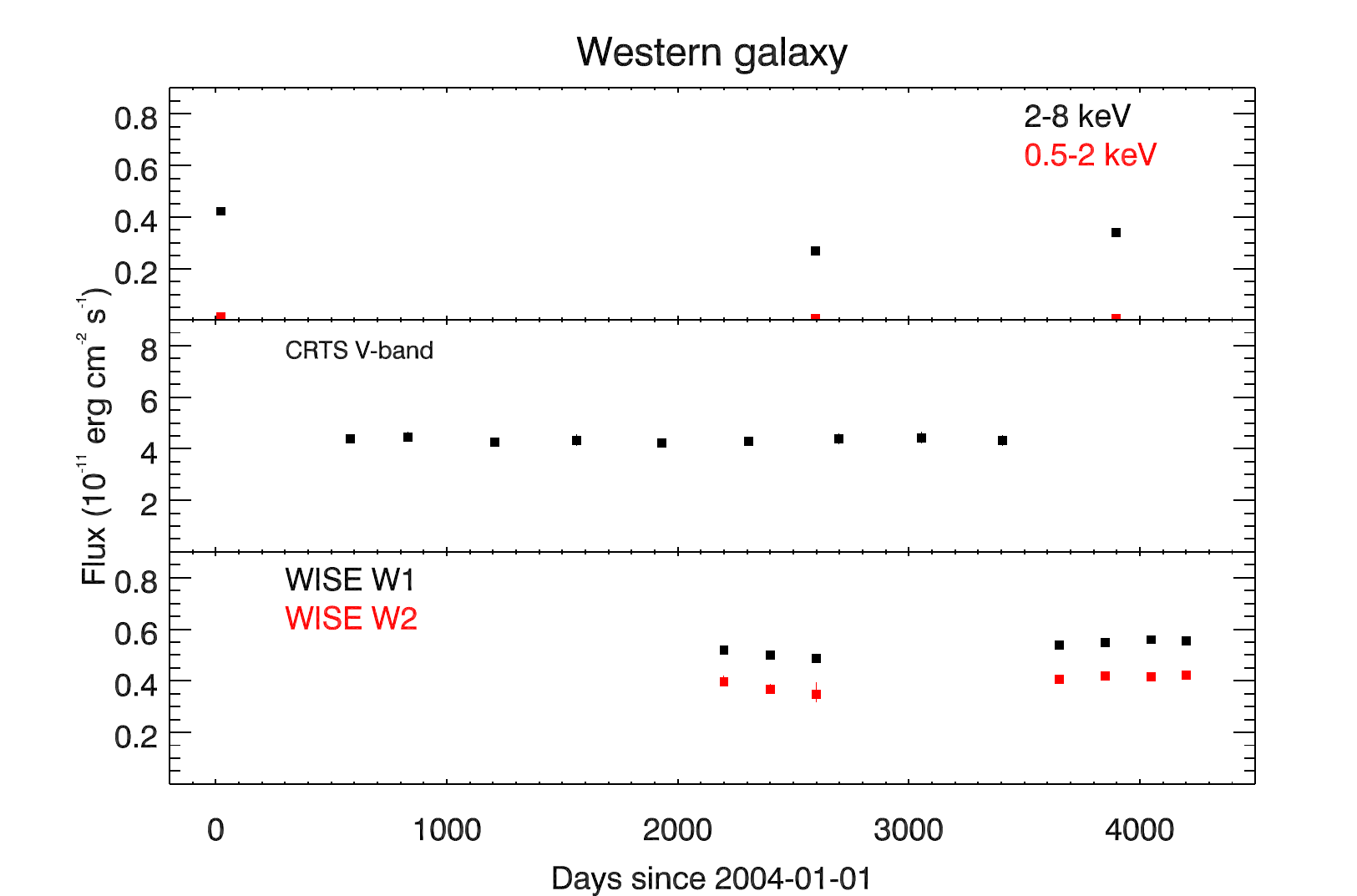}
\caption{Multiband lightcurves of both Eastern (top) and Western (bottom) galaxies covering the period 2004$-$2014. While the X-ray flux from the Eastern galaxy has dropped by a factor of 10 in the X-ray bands, the flux at optical and mid-IR wavelengths has remained constant. No major variations are seen in the flux from the Western galaxy at X-ray, optical or mid-IR wavelengths.}
\label{fig_lightcurves}
\end{center}
\end{figure}

We convert the $V$-band magnitude from CRTS and the {\it WISE} W1 and W2 magnitudes to $\nu F_{\nu}$ fluxes in order to compare to the X-ray data. We plot this multiband lightcurve in Figure \ref{fig_lightcurves}. While the X-ray flux from the Eastern galaxy has dropped by a factor of 10 in the X-ray bands, the flux at optical and mid-IR wavelengths has remained relatively constant over the long base line. No major variations are seen in the flux from the Western galaxy at X-ray, optical or mid-IR wavelengths.

\section{Optical and IR spectroscopy}
\label{sec_spectro}

\begin{figure}
\begin{center}
\includegraphics[width=95mm]{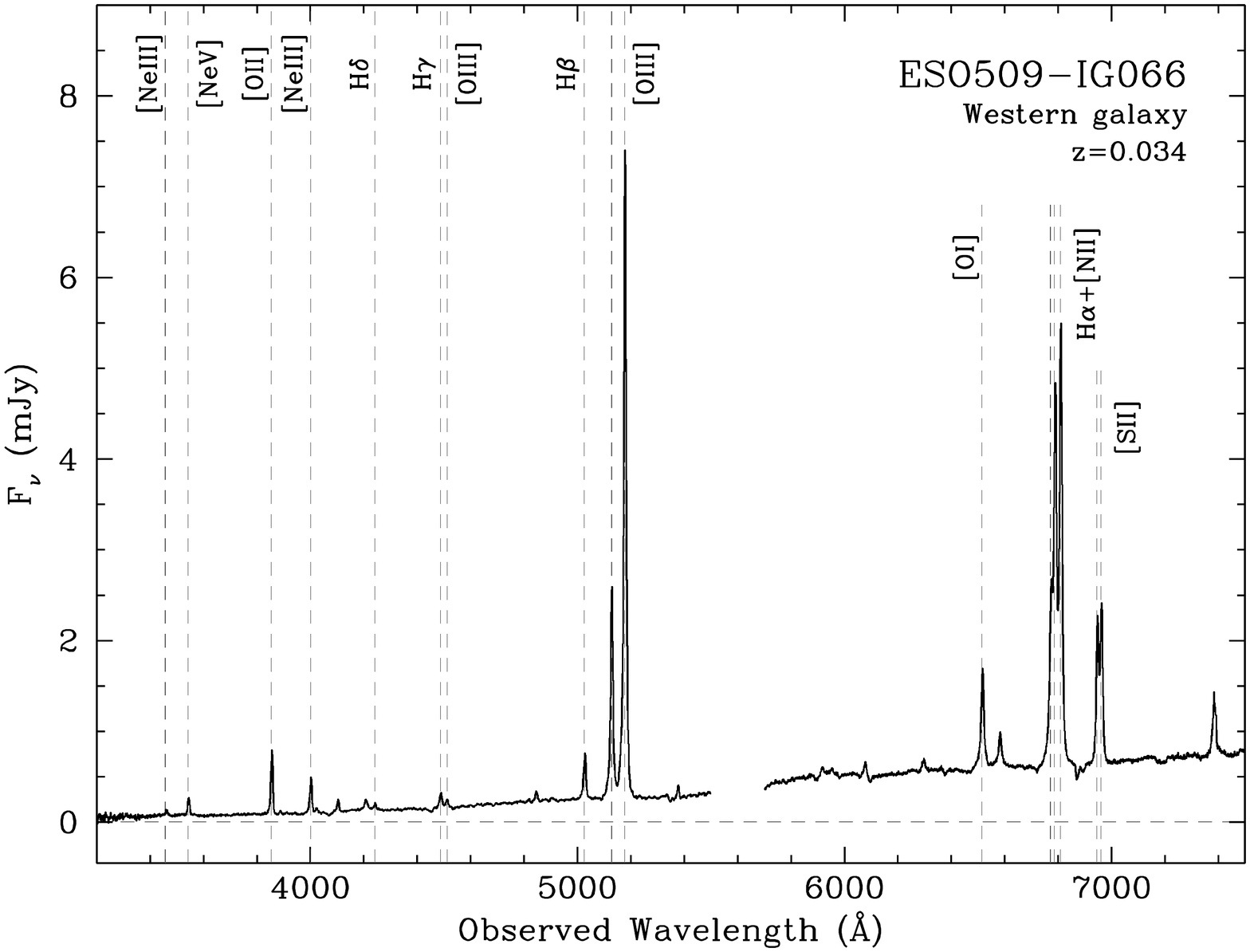}
\includegraphics[width=95mm]{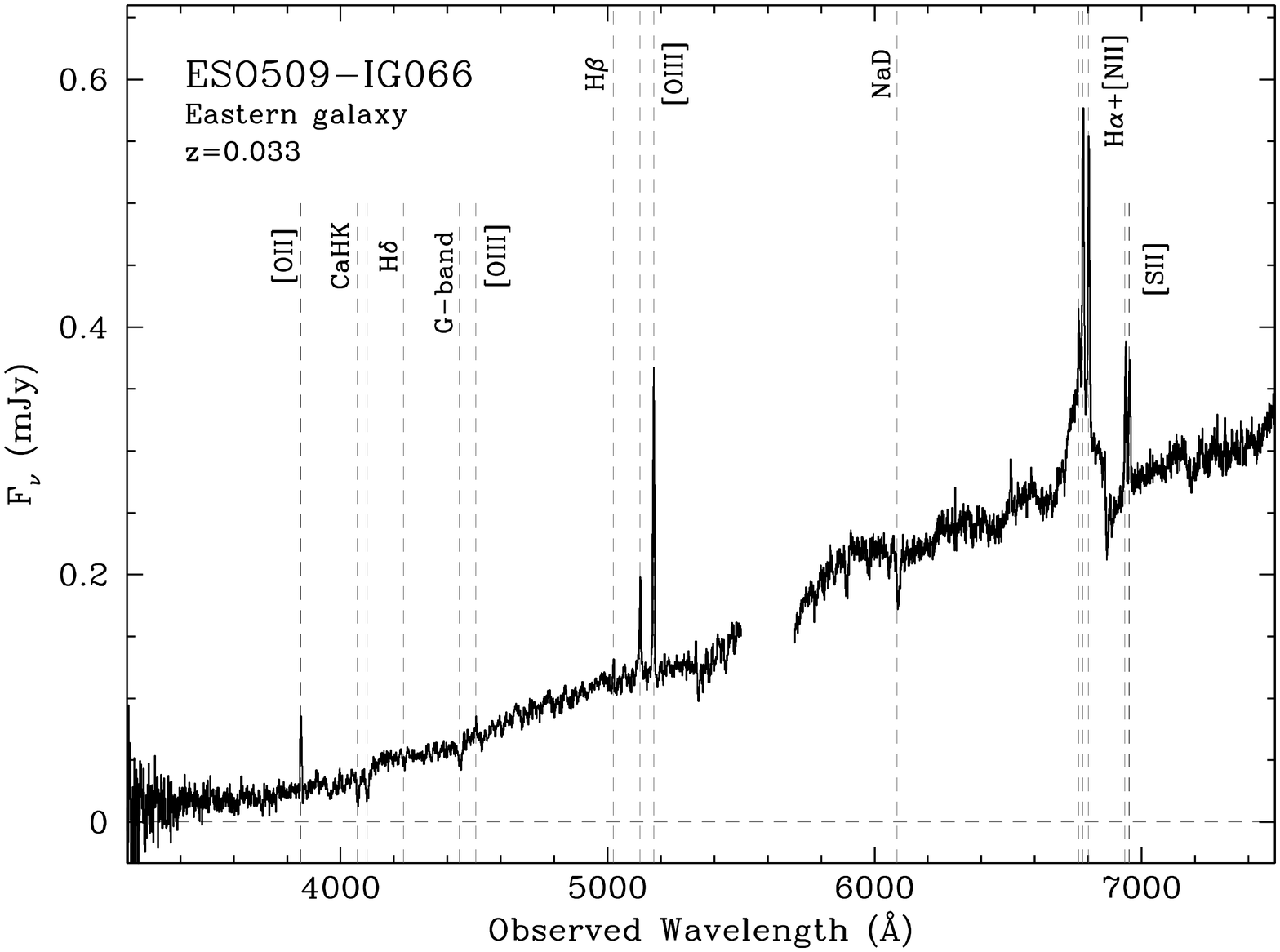}
\caption{Keck/LRIS optical spectra of the nucleus of the Western galaxy (top) showing strong forbidden lines typical of a Seyfert 2 and the Eastern galaxy (bottom) which reveals a broad \ha\ beneath narrow \ha+\nii\ lines. This is the first reported detection of broad optical lines from this galaxy and indicates that the nucleus is not strongly absorbed, arguing against heavy absorption behind the drop in X-ray flux.}
\label{fig_lris}
\end{center}
\end{figure}

In order to gain further insight into the nature of the interaction between the galaxies and the drop in X-ray flux from the Eastern nucleus, we analyzed the optical spectra of the two nuclei. The processed Keck/LRIS spectra are shown in Figure \ref{fig_lris}. We use the penalised PiXel Fitting software \citep[{\tt pPXF},][]{cappellari04} to measure stellar kinematics and the central stellar velocity dispersion with the Indo-U.S. CaT, and MILES empirical stellar library \citep[$3465-9468$\,\AA][]{vazdekis12}.   We fit the residual spectra for emission lines after subtracting the stellar templates with the {\tt PYSPECKIT} software following \citet[][]{berney15} and correct the narrow line ratios (\ha/\hb) assuming an intrinsic ratio of $R=3.1$ and the \citet{cardelli89} reddening curve.  In the case of a \hb\ non-detection, we assume the 3$\sigma$ upper limits for the extinction correction.

\begin{figure*}
\begin{center}
\includegraphics[width=180mm]{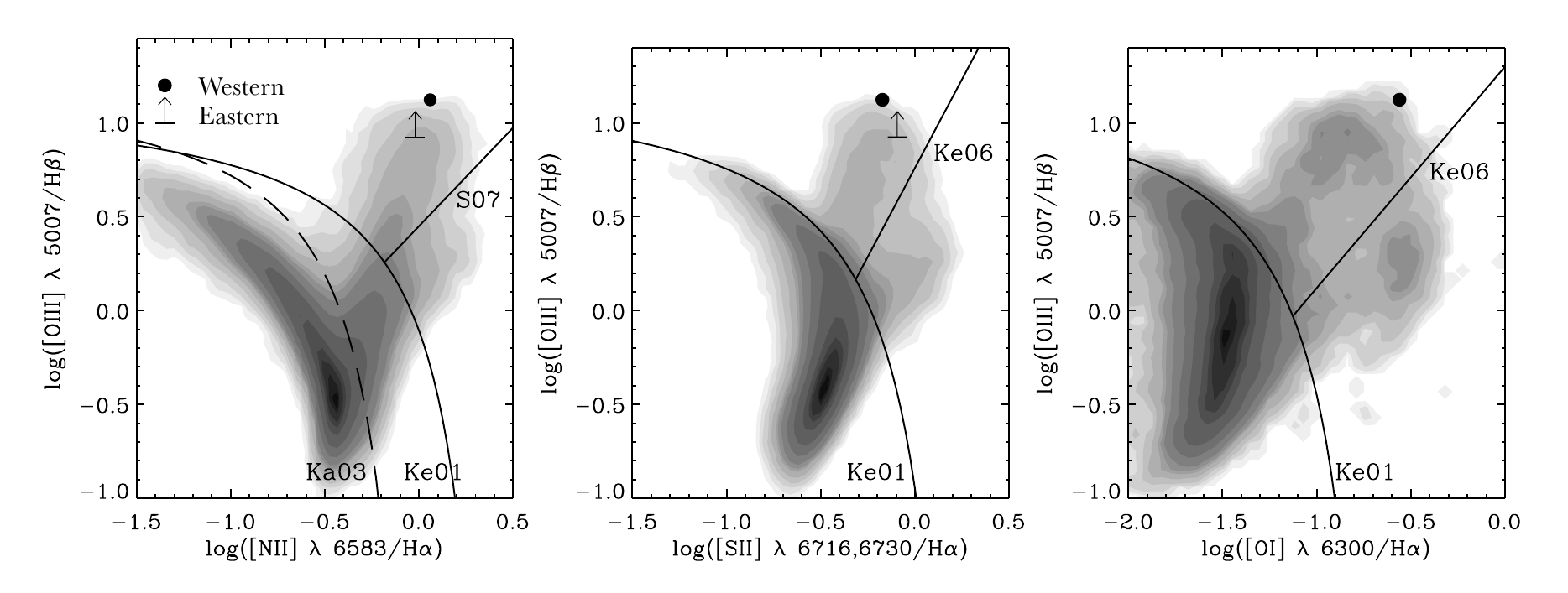}
\caption{ BPT narrow emission-line diagnostic diagrams for the Western (black dot) and Eastern (black bar) nuclei. The solid black curve shows the separation between star forming galaxies, which lie below the curve, and AGN, which lie above the curve, from \cite{kewley01}. The dashed curved line shows the same separation, but from \cite{kauffmann03}. The solid straight line shows the separation between Seyferts, which fall left of the line, and LINERs, which fall to the right of the line, from \cite{kewley06}. The Western nucleus is in the Seyfert section in all three diagrams. The Eastern nucleus only has an upper limit on the \hb\ flux, but the corresponding lower limit of the ratio is in the Seyfert region, thus both galaxies are classified as Seyferts from our data.}
\label{fig_bpt}
\end{center}
\end{figure*}

\begin{figure}
\begin{center}
\includegraphics[width=95mm]{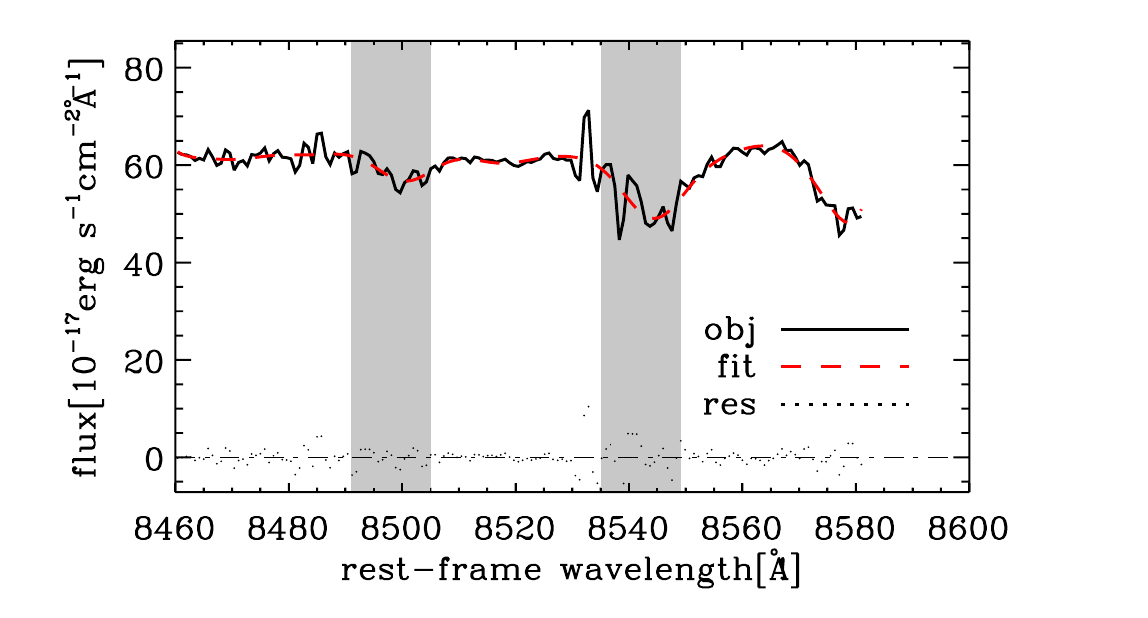}
\caption{Zoom in of the Keck/LRIS optical spectrum of the Western galaxy in the region where the Calcium triplet absorption occurs, which we use to measure the velocity dispersion of the stars.}
\label{fig_catrip}
\end{center}
\end{figure}

The optical spectrum of the Western galaxy exhibits strong forbidden transition lines from \oiii\ and \nii\ and BPT diagnostics confirm that the galaxy hosts a Seyfert 2 nucleus (Figure \ref{fig_bpt}). The Balmer decrement corrected \oiii\ flux is 1.91$\times10^{-13}$ \ergcms. We measure a velocity dispersion of 118$\pm$37 \kms\ in the CaH+K and Mgb region and 124$\pm$27 \kms\ in the Calcium triplet absorption lines. We show the fit to the Calcium triplet lines in Figure \ref{fig_catrip}.

\begin{figure}
\begin{center}
\includegraphics[width=80mm]{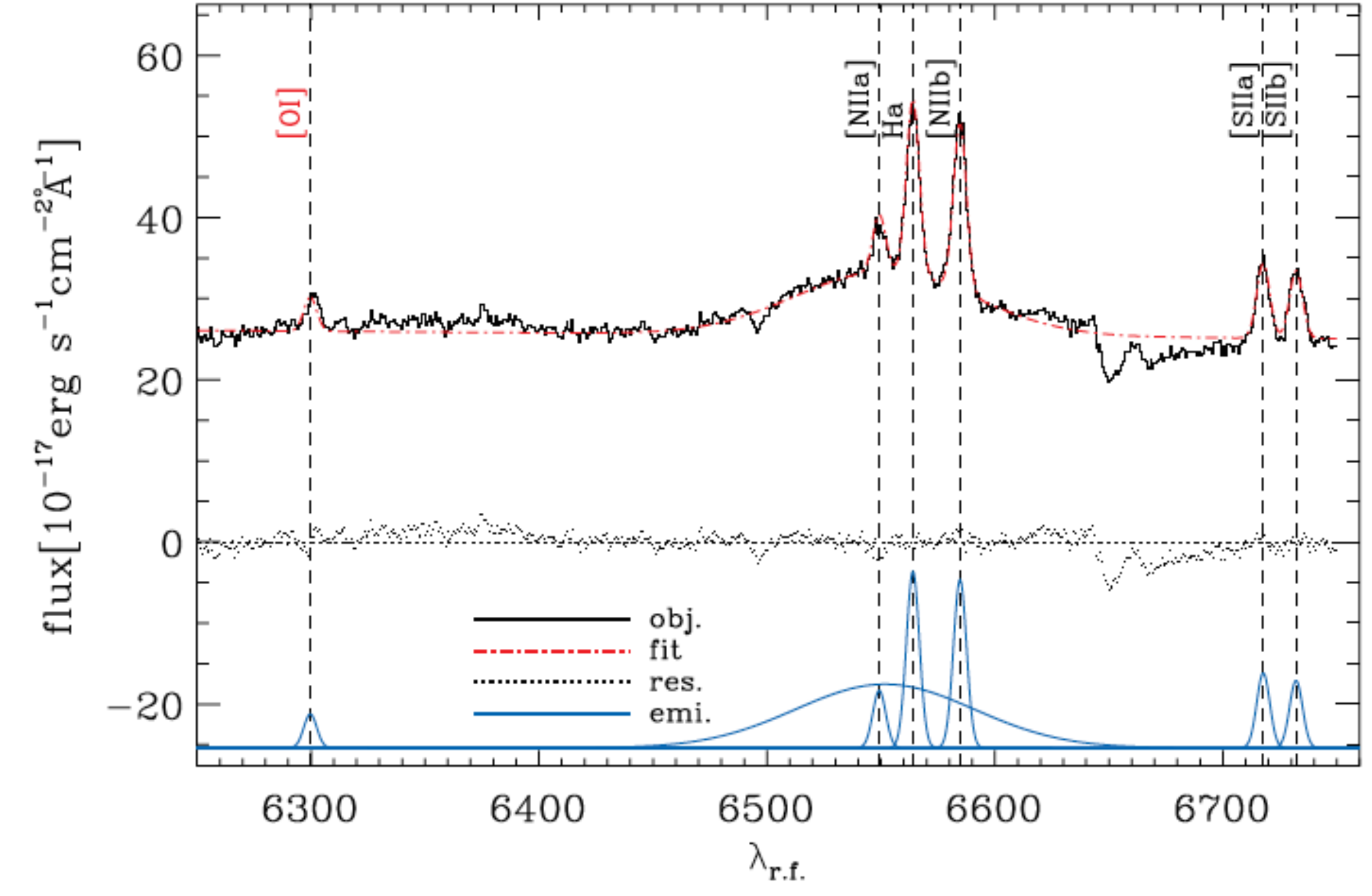}
\caption{Zoom in of the Keck/LRIS optical spectrum of the Eastern galaxy in the region where the broad \ha\ line was detected which shows the spectral decomposition.}
\label{fig_halpha}
\end{center}
\end{figure}

In the LRIS spectrum of the Eastern galaxy a broad \ha\ line is detected with a width of 4226 km\,s$^{-1}$ characteristic of a Seyfert 1 nucleus (Figure \ref{fig_halpha}), however the \hb\ line is very weak, and so would be classified as a Seyfert 1.9 \citep{osterbrock81}. Using an upper limit to the flux of the narrow \hb\ line, we find that the BPT diagnostics also confirm the presence of a Seyfert nucleus in this source (Figure \ref{fig_bpt}). The Eastern galaxy has a velocity dispersion that is consistent with the instrumental resolution ($<$100 \kms) in the CaH+K and Mgb region. However, due to the broad \ha\ line, the Calcium triplet absorption lines are likely contaminated by AGN emission.  

\cite{sekiguchi92} presented optical spectroscopic observations of the two galaxies, taken with the 1.9-m South African Astronomical Observatory, also finding the Western nucleus to be a Seyfert 2. They, however, classify the Eastern nucleus as LINER or \hii\ galaxy. It is unclear if the broad \ha\ line was undetected in their observations or not present at that time when they were made, between 1987 and 1990. Our new detection of broad \ha\ strongly suggests that our view of the Eastern nucleus is largely unobscured. Some reddening may be present in order to explain the non-detection of a broad \hb\ line. We can estimate the amount of reddening from the flux ratio of the \ha\ and \hb\ lines, known as the Balmer decrement. Given the upper limit on the flux of the \hb\ line, the lower limit on the Balmer decrement is 4.7. Assuming an intrinsic value of 3.1, this corresponds to a lower limit on the reddening of $E(B-V)=0.36$, which, assuming the Galactic gas-to-dust ratio corresponds to \nh$\sim10^{21}$ \cmsq. In order to suppress the X-ray flux from the Eastern nucleus such that it is not detected by \nustar, the obscuration must be at least 4 orders of magnitude higher, around $10^{25}$ \cmsq. If this were the case, the implied reddening in the optical means that the broad \ha\ line would not be detectable. A possibility remains, however, that the broad \ha\ line results from scattered light from the nucleus which would not be subjected to the heavy line of sight absorption that the X-rays may be subjected to. 

\begin{figure}
\begin{center}
\includegraphics[width=90mm]{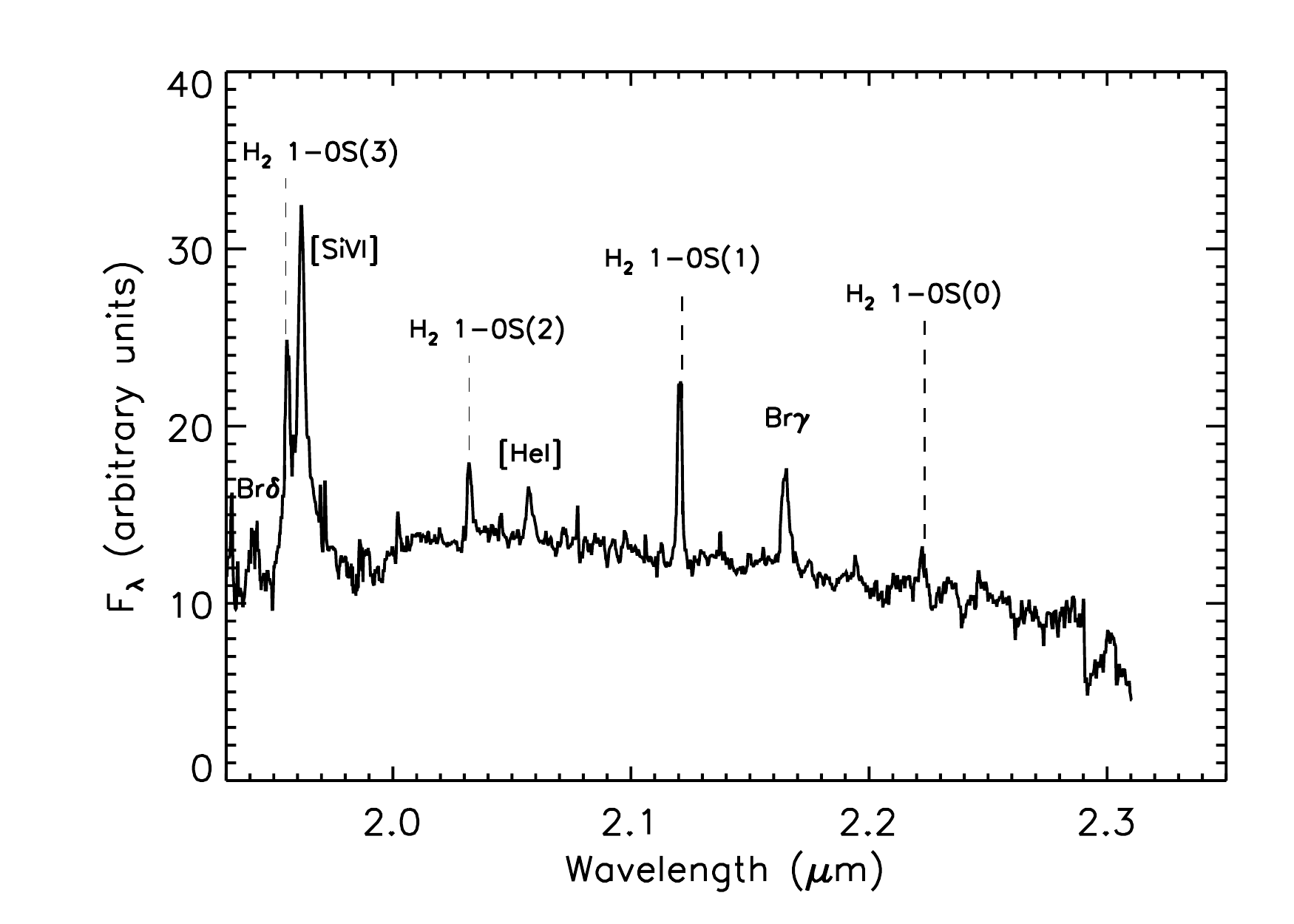}
\caption{Integrated Keck/OSIRIS spectrum of the Western nucleus in the $K$-band (rest frame). The individual spectra were added over an aperture of 0.6\arcsec\ diameter centered at the peak of continuum emission in the near-IR. Several transitions of molecular hydrogen can be seen, where the 2.12\micron\ H$_2$ 1$-$0 S(1) emission line is the strongest. Ionized gas emission is also detected in this galaxy ([Si\,{\sc vi}], Br$\delta$, Br$\gamma$ and [He\,{\sc i}]), confirming the presence of a powerful AGN.}
\label{fig_osiris}
\end{center}
\end{figure}

In addition to the optical spectroscopy of the two nuclei, we also acquired near-IR $K$-band integral field spectroscopy of the inner $\sim$kpc of both nuclei. The integrated spectrum of the Western nucleus (Figure \ref{fig_osiris}) reveals several transitions of molecular hydrogen, Br$\gamma$ and Br$\delta$ from atomic hydrogen, as well as transitions from ionized gas ([Si\,{\sc vi}] and [He\,{\sc i}]). The molecular hydrogen transitions indicate the presence of large amounts of molecular gas while the emission from highly ionized gas confirms the presence of a powerful AGN. The integrated spectrum of the Eastern nucleus is however featureless, with no emission lines detected.

\begin{figure}
\begin{center}
\includegraphics[width=90mm]{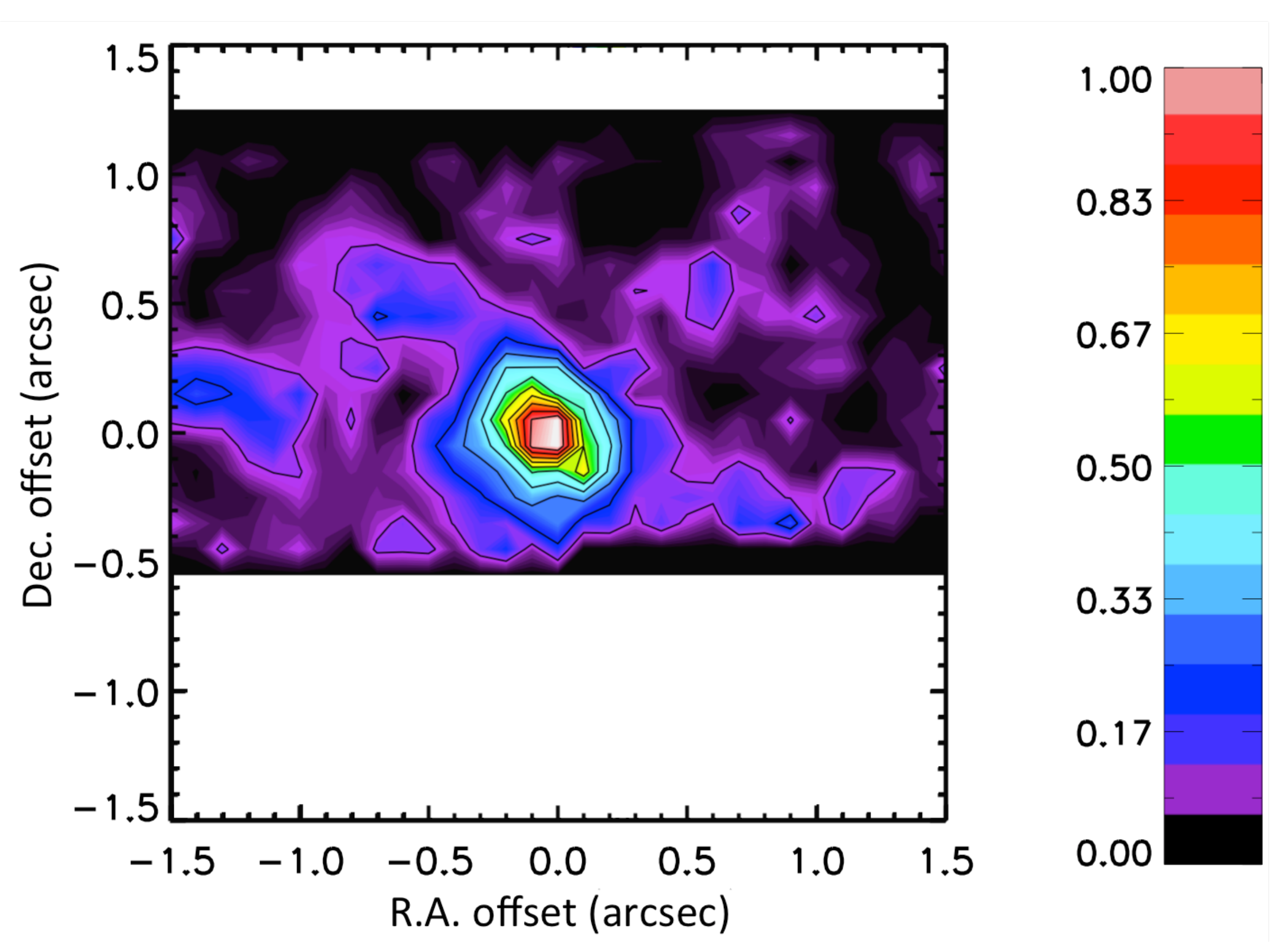}
\includegraphics[width=90mm]{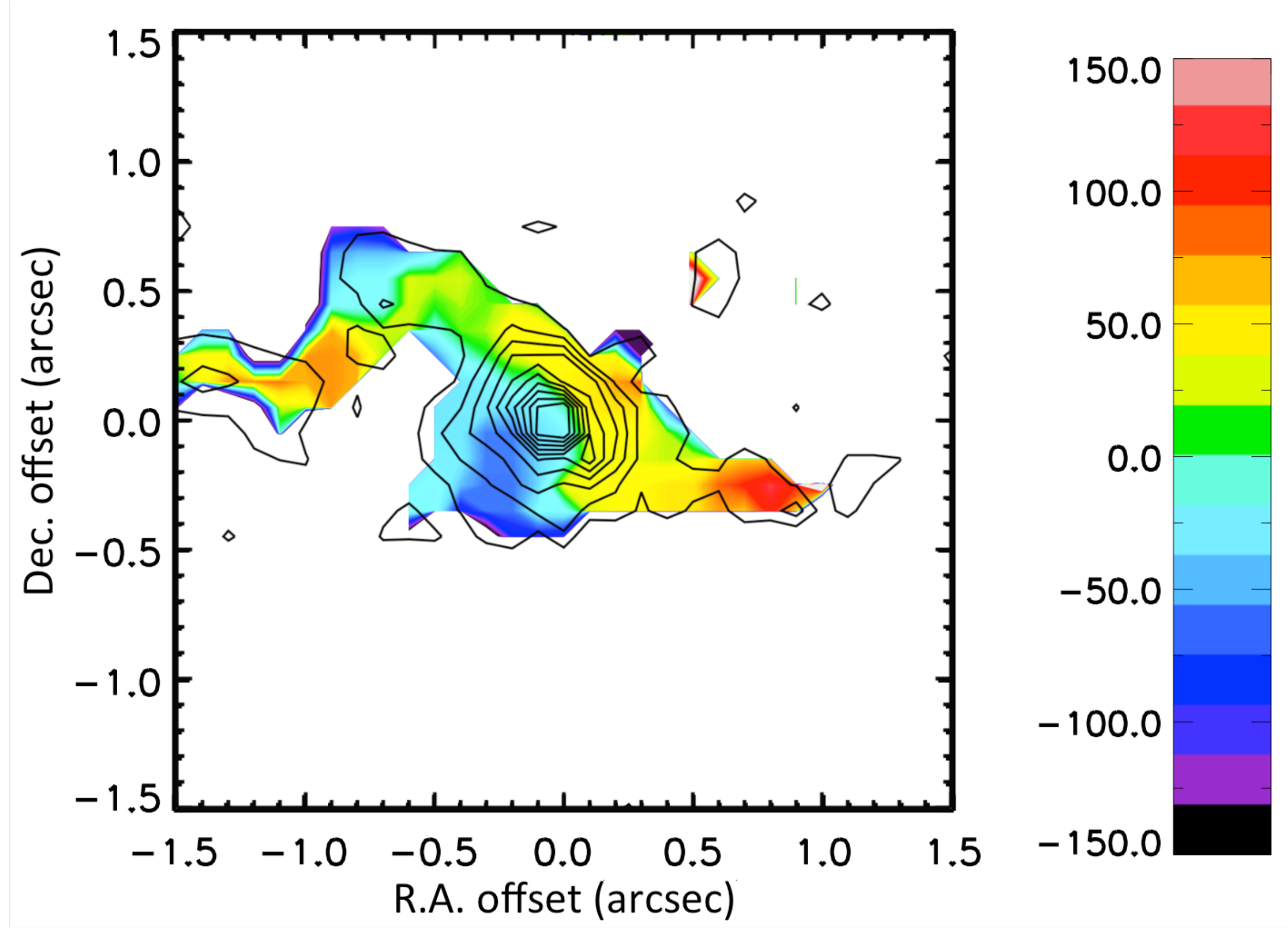}
\includegraphics[width=90mm]{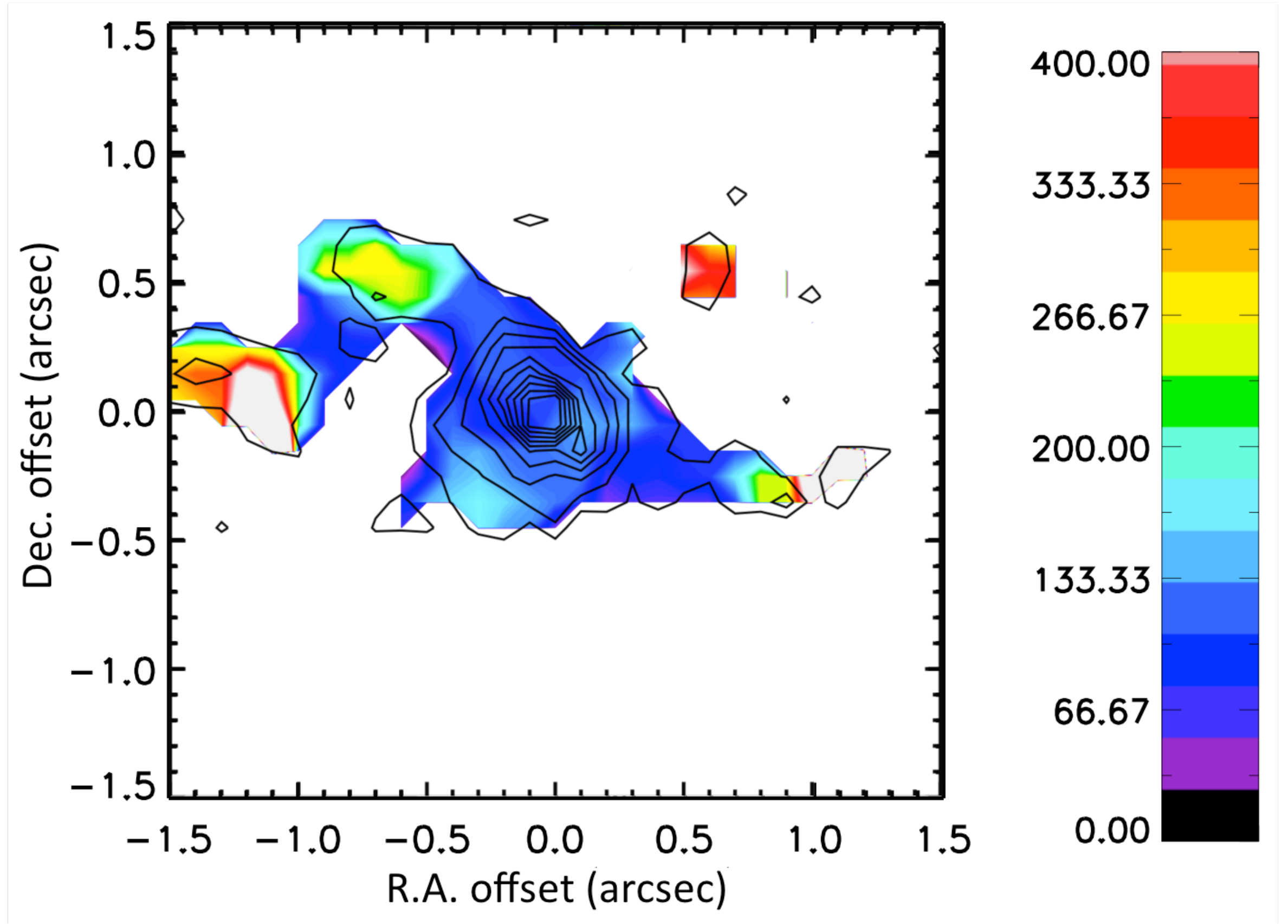}
\caption{Maps of the H$_2$ 1$-$0 S(1) flux (top), velocity (middle) and velocity dispersion (bottom) in the nucleus of the Western galaxy acquired from Keck/OSIRIS near-IR integral-field spectroscopy observations. The color scale is in km s$^{-1}$ and the angular scale is 700 pc/\arcsec\ at the redshift of the system and our assumed cosmology. The central 200$-$300 pc shows ordered rotation (PA of the kinematic major axis $\sim$135\degree), whereas the gas at $\sim$1 kpc to the east of the nucleus appears to be systematically redshifted (top). The same region of gas also shows a high (455 km s$^{-1}$) velocity dispersion, pointing towards a physical interaction between the galaxies (bottom). The contours delineate the molecular gas morphology and are normalized to the peak of emission. Each contour represents a change in flux of 10\%.}
\label{fig_maps}
\end{center}
\end{figure}

We use the strong H$_2$ 1$-$0 S(1) emission line at 2.12 \micron\ to map the velocity of the gas within the inner $\sim$kpc of the Western galaxy, fitting it with a single Gaussian for each pixel in the field of view. Figure \ref{fig_maps} presents the flux, velocity and the velocity dispersion inferred from these measurements. We find that within the inner 200$-$300 pc of the galaxy, the gas rotates in an ordered fashion, with a systematic velocity towards us to the west of the nucleus, and a systematic velocity away from us to the east of the nucleus. The velocity dispersion is also low ($<100$ km s$^{-1}$). However, at $\sim1$ kpc to the east of the nucleus, a region of gas appears to have motion that is redshifted towards us, opposite to the direction of ordered rotation in that region. This region also shows very high velocity dispersion of 455 km s$^{-1}$ (typical error 10--20 km s$^{-1}$). A high dispersion is an indication of shocks and perturbed kinematics. These usually correspond to outflows, but they are also associated with inflows, particularly from merger processes \citep[][]{medling15,msanchez16}. We have mapped the gas outflow from the Western AGN using the high-ionisation [Si\,{\sc vi}] line which shows a different morphology from the molecular gas, orientated in the north-south direction. We therefore find it unlikely that the perturbed molecular gas in the east is caused by an AGN outflow. Since this galaxy appears to be interacting with its neighbor to the east, we interpret these observations as signatures of an inflow of gas caused by a physical interaction between the galaxies. While a region of high velocity dispersion is also seen to the north west of the nucleus, the signal to noise is low and has significantly lower velocity dispersion (300 km s$^{-1}$) than the region to the east. Furthermore, there is no evidence that the velocity is systematically different to the ordered rotation seen in the rest of the nucleus.

Therefore we conclude that while the nuclei are separated by $\sim11$ kpc, the effect of the interaction is seen on the gas within the inner $\sim1$ kpc of the Western galaxy. This clearly shows that galaxy interactions like the one is ESO 509-IG066 can have significant impact on the motion of gas within the nuclei and the feeding of the central SMBH.

\section{Discussion}
\label{sec_disc}

One of the notable features of the X-ray observations spanning 12 years is the drop, by a factor of 10, of the flux of the Eastern source. There are two possible straightforward explanations for this observed drop in X-ray flux. The decrease in flux could be caused by an increase in the column density of the absorber. A cloud of gas and dust might be passing in front of the nucleus along the line of sight obscuring the source \citep[e.g. NGC 1365,][]{risaliti09, rivers15}. Provided the column density is extremely high (above 10$^{25}$ \cmsq), the only received X-ray radiation would be that escaping through gaps between clouds, assuming that the covering fraction is not 100\%, or light that has been scattered into our line of sight. The spectrum would then resemble results obtained from the 2011 \chandra\ observation, being lower in flux with approximately the same spectral shape. An extreme \nh\  would be required so that even emission above 10 keV is suppressed by Compton scattering since the Eastern source is not visible in the \nustar\ image (Figure \ref{Xray_image}). With an angular separation of 16\arcsec, the nuclei are far enough apart to be distinguishable with \nustar, whose PSF has a 18\arcsec\ FWHM \citep{harrison13,madsen15}. The X-ray emission in the \nustar\ image peaks strongly at the position of the Western source, with no indication of the Eastern one. Furthermore, there is no evidence for Fe-K$\alpha$ emission in the \chandra\ spectrum of the source. In AGN, significant obscuration is usually, but not always, associated with Fe-K$\alpha$ fluorescence emission.  The absence of an Fe-K line suggests absorption is not responsible for the flux decrease. 

The second possibility is that the intrinsic X-ray luminosity of the AGN itself decreased by a factor of at least 10 over the past 10 years, due to a decrease in coronal activity which could have been caused by a drop in mass accretion rate. This would explain the spectral shape seen in the 2011 \chandra\ observation, which is fitted well by a model with similar physical parameters such as column density and the power-law slope. Assuming a similar luminosity of the nucleus in 2014 and 2011, this scenario also agrees with the non-detection of the AGN by \nustar\ and its very faint detection by \swiftxrt. The Eastern nucleus was weak in the hard X-ray band in 2004 with respect to the Western source so would be undetectable by \nustar\ above 3 keV after a decrease by a factor of 10. Additional evidence in favor of the drop in accretion rate comes from the optical spectrum, which reveals a weak, but significantly detected broad \ha\ line from the nucleus of the Eastern galaxy, which must come from close to the central engine. If the dimming were due to obscuration, it would require an extremely low dust-to-gas ratio for the X-ray flux to have undergone such suppression, while the \ha\ line remains visible, although the possibility still exists that the \ha\ line may be scattered light.

However, no emission lines were detected in the NIR from the Eastern nucleus. This implies that there are not enough ionizing photons to produce emission lines of ionized gas in the near-IR (like Br-$\gamma$, [He\,{\sc i}] or [Si\,{\sc vi}] seen from the Western nucleus). Also, the lack of molecular hydrogen indicates that there is not sufficient gas to maintain the active nucleus. This is consistent with our interpretation that the accretion rate of the eastern nucleus has dropped.

We discuss the scenario that the merging of the two galaxies is directly linked to the change in accretion rate of the Eastern AGN. Firstly, galaxy merger simulations predict large fluctuations in black hole accretion rate during the final stages of a merger \citep[e.g.][]{vanwassenhove12,gabor15}. Although the time resolution of most simulations ($\sim10^3$ years) is much longer than our observational time scale, results on much shorter time scales ($\sim10$ years) also reveal similar accretion rate fluctuations (J. Gabor, private communication). It should be noted, however, that fluctuations are also predicted from simulations of isolated AGN \citep[e.g.][]{novak11} and have been observed as well \citep{lamassa15}, although this could be related to a tidal disruption event \citep{merloni15}. Secondly, the motion of the gas in the central $\sim$kpc of the Western galaxy as revealed by integral field spectroscopy is highly suggestive that the galaxy merger is directly affecting the kinematics of the gas within the nuclear region, providing a direct observational link between the galaxy merger and the change in mass accretion rate on to the black hole.

To better place this AGN pair in the context of galaxy simulations, we estimate the masses of the central SMBHs. For the Eastern galaxy, a broad \ha\ line was detected, which we use for the \mbh\ estimation. \cite{greene05} presented a method for estimating the black hole mass from the width and luminosity of the \ha\ line. From their equation 6, given that we measure a width of 4226 \kms\ and a luminosity of 2.1$\times10^{40}$ \ergs, we obtain \mbh$\approx4.6\times10^{6}$ \msol. Since no broad line was detected in the Western galaxy, we use the velocity dispersion of the stars in the center of the galaxy to estimate the black hole mass. Using the \mbh$-\sigma_{*}$ relation from \cite{kormendy13} and the Calcium triplet measurement implies a black hole mass of \mbh$=3.8_{-2.6}^{+5.1}\times10^{7}$ \msol\ for the Western galaxy. This then implies that the black hole mass ratios of the two galaxies is 10:1, which is rather larger than the 4:1 or 2:1 \mbh\ ratios considered in recent simulations by \cite{gabor15}.

It is interesting that the AGN in this system with the smallest black hole mass has exhibited the greatest X-ray variability since it is well known that the variability timescale correlates with black hole mass. I.e. The variability timescale increases with \mbh\ \citep[e.g.][]{papadakis04}. However, these timescales are much shorter ($\sim10^4$s) than the timescale of the drop in X-ray flux we have observed.

An observational signature for the drop in \lamedd\ in the X-ray spectrum of the Eastern source is expected since there is a known correlation between \lamedd\ and $\Gamma$ \citep[e.g.][]{shemmer06,risaliti09,brightman13,brightman16b}. During the 2004 \xmm\ observation, the Eastern source had an absorption-corrected \lx\ of 7.5$\times10^{42}$ \ergs. Applying a bolometric correction of 10 \citep{lusso12} implies \lbol=7.5$\times10^{43}$ \ergs, which in turn yields \lamedd=0.12 for our \mbh\ estimate. At the time of the \chandra\ observation, \lx\ had reduced by a factor of 10, meaning a decline in \lamedd\ by the same factor. For the observations in 2014, the X-ray emission from the Eastern nucleus was undetectable, thus \lamedd\ $\lesssim0.01$ at that time. From \cite{brightman13}, $\Gamma=(0.32\pm0.05)$log$_{10}$\lamedd$+(2.27\pm0.06)$, thus we would expect $\Delta\Gamma\approx-0.3$ for $\Delta$log$_{10}$\lamedd$\approx-1$. For our analysis presented in Section \ref{sec_analysis}, we tie the $\Gamma$ values to each other for both flux levels. If we perform the same analysis, but with the $\Gamma$ parameter not linked between the observations, we obtain $\Gamma=1.49\pm0.12$ for the \xmm\ observation and $\Gamma=1.53^{+0.62}_{-0.57}$ for the \chandra\ observation. The uncertainties are therefore too large to constrain $\Delta\Gamma$ at the requisite level.

\section{Summary and Conclusions}
\label{sec_conc}

We conducted a multi-wavelength analysis of the galaxy pair ESO~509-IG066 using X-ray, optical, near-IR and mid-IR data taken between 2004 and 2016. The pair of galaxies, located at a distance of 150 Mpc with a projected separation of 10.9 kpc, were both reported to host an AGN of \lx$\sim10^{43}$ \ergs\ by G05 using \xmm\ data. In an analysis of all available data, we found that since the \xmm\ observation in 2004, the Eastern nucleus has shown a strong decrease in X-ray flux revealed first by a \chandra\ observation in 2011. The galaxy remained at this level or lower during a joint \nustar\ and \swiftxrt\ observation in 2014. The X-ray emission from the Western source remained relatively constant during this period. Although the 16\arcsec\ angular separation of the galaxy pair causes significant overlap given the \nustar\ PSF, there is no evidence for the Eastern source in the \nustar\ image from 2014. This argues against a rise in obscuration behind the drop in X-ray flux, unless it is extreme. New Keck/LRIS optical spectroscopy taken after the drop in X-ray flux reveals a broad component to the \ha\ line from the Eastern nucleus, which also strongly argues against heavy obscuration. We therefore conclude that the AGN has dropped intrinsically in luminosity, most likely due to a decrease in mass accretion rate. From AO-assisted near-infrared integral-field spectroscopy, we find that the kinematics of the gas close to the Western nucleus show evidence that the galaxy merger is having a direct effect close in to the black hole, providing an observational link between the galaxy merger and the mass accretion rate on to the black hole.

\acknowledgments

We thank the referee for their constructive input on our manuscript. We also thank Mislav Balokovi\'{c} for introducing us to these interesting galaxies and Jared Gabor for useful discussion. RJA was supported by FONDECYT grant number 1151408. AM acknowledges support from the ASI/INAF grant I/037/12/0-011/13. CR acknowledges financial support from the CONICYT-Chile grants ``EMBIGGEN" Anillo ACT1101, FONDECYT 1141218, Basal-CATA PFB--06/2007 and from the China-CONICYT fund. This work was supported under NASA Contract No. NNG08FD60C, and made use of data from the {\it NuSTAR} mission, a project led by the California Institute of Technology, managed by the Jet Propulsion Laboratory, and funded by the National Aeronautics and Space Administration. We thank the {\it NuSTAR} Operations, Software and Calibration teams for support with the execution and analysis of these observations.  This research has made use of the {\it NuSTAR} Data Analysis Software (NuSTARDAS) jointly developed by the ASI Science Data Center (ASDC, Italy) and the California Institute of Technology (USA). The work presented here was also based on observations obtained with \xmm, an ESA science mission with instruments and contributions directly funded by ESA Member States and NASA. This publication makes use of data products from the Wide-field Infrared Survey Explorer, which is a joint project of the University of California, Los Angeles, and the Jet Propulsion Laboratory/California Institute of Technology, funded by the National Aeronautics and Space Administration. The data presented herein were obtained at the W.M. Keck Observatory, which is operated as a scientific partnership among the California Institute of Technology, the University of California and the National Aeronautics and Space Administration. The Observatory was made possible by the generous financial support of the W.M. Keck Foundation. The authors wish to recognize and acknowledge the very significant cultural role and reverence that the summit of Mauna Kea has always had within the indigenous Hawaiian community.  We are most fortunate to have the opportunity to conduct observations from this mountain. This research has also made use of data and software provided by the High Energy Astrophysics Science Archive Research Center (HEASARC), which is a service of the Astrophysics Science Division at NASA/GSFC and the High Energy Astrophysics Division of the Smithsonian Astrophysical Observatory. Furthermore, this research has made use of the NASA/IPAC Extragalactic Database (NED) which is operated by the Jet Propulsion Laboratory, California Institute of Technology, under contract with the National Aeronautics and Space Administration. AM acknowledges support from the ASI/INAF grant I/037/12/0-011/13, CR acknowledges financial support from the CONICYT-Chile grants ``EMBIGGEN" Anillo ACT1101, FONDECYT 1141218, Basal-CATA PFB--06/2007 and from the China-CONICYT fund.

{\it Facilities:} \facility{\nustar, \xmm\ pn, \chandra, \swift\ (XRT) {\it WISE}, {\it NEOWISE}, Keck/LRIS, Keck/OSIRIS}

\bibliography{bibdesk}

\begin{thebibliography}{}

\bibitem[\protect\astroncite{{Alonso} et~al.}{2007}]{alonso07}
{Alonso}, M.~S., {Lambas}, D.~G., {Tissera}, P., \& {Coldwell}, G.,  2007,
  \mnras, 375, 1017

\bibitem[\protect\astroncite{{Arnaud}}{1996}]{arnaud96}
{Arnaud}, K.~A.,  1996,
\newblock in Astronomical Data Analysis Software and Systems V, ed. G.~H.
  {Jacoby}, J. {Barnes}, Vol. 101, 17

\bibitem[\protect\astroncite{{Baumgartner} et~al.}{2013}]{baumgartner13}
{Baumgartner}, W.~H., {Tueller}, J., {Markwardt}, C.~B., {Skinner}, G.~K.,
  {Barthelmy}, S., {Mushotzky}, R.~F., {Evans}, P.~A., \& {Gehrels}, N.,  2013,
  \apjs, 207, 19

\bibitem[\protect\astroncite{{Berney} et~al.}{2015}]{berney15}
{Berney}, S., et~al., 2015, \mnras, 454, 3622

\bibitem[\protect\astroncite{{Brightman} et~al.}{2016}]{brightman16b}
{Brightman}, M., et~al., 2016, \apj, 826, 93

\bibitem[\protect\astroncite{{Brightman} et~al.}{2013}]{brightman13}
{Brightman}, M., {Silverman}, J.~D., {Mainieri}, V., {Ueda}, Y., {Schramm}, M.,
  {Matsuoka}, K., {Nagao}, T., \& {Steinhardt}, C.,  2013, \mnras, 433, 2485

\bibitem[\protect\astroncite{{Burrows} et~al.}{2005}]{burrows05}
{Burrows}, D.~N., et~al., 2005, Space Science Reviews, 120, 165

\bibitem[\protect\astroncite{{Cappellari} \& {Emsellem}}{2004}]{cappellari04}
{Cappellari}, M., \& {Emsellem}, E.,  2004, \pasp, 116, 138

\bibitem[\protect\astroncite{{Cardelli}, {Clayton} \&
  {Mathis}}{1989}]{cardelli89}
{Cardelli}, J.~A., {Clayton}, G.~C., \& {Mathis}, J.~S.,  1989, \apj, 345, 245

\bibitem[\protect\astroncite{{Cash}}{1979}]{cash79}
{Cash}, W.,  1979, \apj, 228, 939

\bibitem[\protect\astroncite{{Cusumano} et~al.}{2010}]{cusumano10}
{Cusumano}, G., et~al., 2010, \aap, 524, A64+

\bibitem[\protect\astroncite{{Di Matteo}, {Springel} \&
  {Hernquist}}{2005}]{dimatteo05}
{Di Matteo}, T., {Springel}, V., \& {Hernquist}, L.,  2005, \nat, 433, 604

\bibitem[\protect\astroncite{{Drake} et~al.}{2009}]{drake09}
{Drake}, A.~J., et~al., 2009, \apj, 696, 870

\bibitem[\protect\astroncite{{Ellison} et~al.}{2011}]{ellison11}
{Ellison}, S.~L., {Patton}, D.~R., {Mendel}, J.~T., \& {Scudder}, J.~M.,  2011,
  \mnras, 418, 2043

\bibitem[\protect\astroncite{{Gabor} et~al.}{2015}]{gabor15}
{Gabor}, J.~M., {Capelo}, P.~R., {Volonteri}, M., {Bournaud}, F., {Bellovary},
  J., {Governato}, F., \& {Quinn}, T.,  2015, ArXiv e-prints

\bibitem[\protect\astroncite{{Gandhi} et~al.}{2017}]{gandhi17}
{Gandhi}, P., et~al., 2017, \mnras, 467, 4606

\bibitem[\protect\astroncite{{Gehrels} et~al.}{2004}]{gehrels04}
{Gehrels}, N., et~al., 2004, \apj, 611, 1005

\bibitem[\protect\astroncite{{Greene} \& {Ho}}{2005}]{greene05}
{Greene}, J.~E., \& {Ho}, L.~C.,  2005, \apj, 630, 122

\bibitem[\protect\astroncite{{Guainazzi} et~al.}{2005}]{guainazzi05}
{Guainazzi}, M., {Piconcelli}, E., {Jim{\'e}nez-Bail{\'o}n}, E., \& {Matt}, G.,
   2005, \aap, 429, L9

\bibitem[\protect\astroncite{{Harrison} et~al.}{2013}]{harrison13}
{Harrison}, F.~A., {Craig}, W.~W., {Christensen}, F.~E., {Hailey}, C.~J., \&
  {Zhang}, W.~W.,  2013, \apj, 770, 103

\bibitem[\protect\astroncite{{Hernquist}}{1989}]{hernquist89}
{Hernquist}, L.,  1989, \nat, 340, 687

\bibitem[\protect\astroncite{{Hiroi} et~al.}{2011}]{hiroi11}
{Hiroi}, K., et~al., 2011, \pasj, 63, 677

\bibitem[\protect\astroncite{{Hopkins} et~al.}{2005}]{hopkins05}
{Hopkins}, P.~F., {Hernquist}, L., {Cox}, T.~J., {Di Matteo}, T., {Martini},
  P., {Robertson}, B., \& {Springel}, V.,  2005, \apj, 630, 705

\bibitem[\protect\astroncite{{Hopkins} et~al.}{2006}]{hopkins06}
{Hopkins}, P.~F., {Hernquist}, L., {Cox}, T.~J., {Di Matteo}, T., {Robertson},
  B., \& {Springel}, V.,  2006, \apjs, 163, 1

\bibitem[\protect\astroncite{{Jansen} et~al.}{2001}]{jansen01}
{Jansen}, F., et~al., 2001, \aap, 365, L1

\bibitem[\protect\astroncite{{Kalberla} et~al.}{2005}]{kalberla05}
{Kalberla}, P.~M.~W., {Burton}, W.~B., {Hartmann}, D., {Arnal}, E.~M.,
  {Bajaja}, E., {Morras}, R., \& {P{\"o}ppel}, W.~G.~L.,  2005, \aap, 440, 775

\bibitem[\protect\astroncite{{Kauffmann} et~al.}{2003}]{kauffmann03}
{Kauffmann}, G., et~al., 2003, \mnras, 346, 1055

\bibitem[\protect\astroncite{{Kewley} et~al.}{2006}]{kewley06}
{Kewley}, L.~J., {Groves}, B., {Kauffmann}, G., \& {Heckman}, T.,  2006,
  \mnras, 372, 961

\bibitem[\protect\astroncite{{Kewley} et~al.}{2001}]{kewley01}
{Kewley}, L.~J., {Heisler}, C.~A., {Dopita}, M.~A., \& {Lumsden}, S.,  2001,
  ApJS, 132, 37

\bibitem[\protect\astroncite{{Kocevski} et~al.}{2015}]{kocevski15}
{Kocevski}, D.~D., et~al., 2015, \apj, 814, 104

\bibitem[\protect\astroncite{{Kormendy} \& {Ho}}{2013}]{kormendy13}
{Kormendy}, J., \& {Ho}, L.~C.,  2013, \araa, 51, 511

\bibitem[\protect\astroncite{{Koss} et~al.}{2012}]{koss12}
{Koss}, M., {Mushotzky}, R., {Treister}, E., {Veilleux}, S., {Vasudevan}, R.,
  \& {Trippe}, M.,  2012, \apjl, 746, L22

\bibitem[\protect\astroncite{{LaMassa} et~al.}{2015}]{lamassa15}
{LaMassa}, S.~M., et~al., 2015, \apj, 800, 144

\bibitem[\protect\astroncite{{Larkin} et~al.}{2006}]{larkin06}
{Larkin}, J., et~al., 2006, \nar, 50, 362

\bibitem[\protect\astroncite{{Lusso} et~al.}{2012}]{lusso12}
{Lusso}, E., et~al., 2012, \mnras, 425, 623

\bibitem[\protect\astroncite{{Madsen} et~al.}{2015}]{madsen15}
{Madsen}, K.~K., et~al., 2015, \apjs, 220, 8

\bibitem[\protect\astroncite{{Magdziarz} \& {Zdziarski}}{1995}]{magdziarz95}
{Magdziarz}, P., \& {Zdziarski}, A.~A.,  1995, \mnras, 273, 837

\bibitem[\protect\astroncite{{Mainzer} et~al.}{2011}]{mainzer11}
{Mainzer}, A., et~al., 2011, \apj, 731, 53

\bibitem[\protect\astroncite{{Malkan}, {Gorjian} \& {Tam}}{1998}]{malkan98}
{Malkan}, M.~A., {Gorjian}, V., \& {Tam}, R.,  1998, \apjs, 117, 25

\bibitem[\protect\astroncite{{Medling} et~al.}{2015}]{medling15}
{Medling}, A.~M., et~al., 2015, \mnras, 448, 2301

\bibitem[\protect\astroncite{{Merloni} et~al.}{2015}]{merloni15}
{Merloni}, A., et~al., 2015, \mnras, 452, 69

\bibitem[\protect\astroncite{{M{\"u}ller-S{\'a}nchez}
  et~al.}{2016}]{msanchez16}
{M{\"u}ller-S{\'a}nchez}, F., {Comerford}, J., {Stern}, D., \& {Harrison},
  F.~A.,  2016, \apj, 830, 50

\bibitem[\protect\astroncite{{Novak}, {Ostriker} \& {Ciotti}}{2011}]{novak11}
{Novak}, G.~S., {Ostriker}, J.~P., \& {Ciotti}, L.,  2011, \apj, 737, 26

\bibitem[\protect\astroncite{{Oke} et~al.}{1995}]{oke95}
{Oke}, J.~B., et~al., 1995, \pasp, 107, 375

\bibitem[\protect\astroncite{{Osterbrock}}{1981}]{osterbrock81}
{Osterbrock}, D.~E.,  1981, \apj, 249, 462

\bibitem[\protect\astroncite{{Papadakis}}{2004}]{papadakis04}
{Papadakis}, I.~E.,  2004, \mnras, 348, 207

\bibitem[\protect\astroncite{{Planck Collaboration} et~al.}{2015}]{planck15}
{Planck Collaboration}et~al., 2015, ArXiv e-prints

\bibitem[\protect\astroncite{{Ricci} et~al.}{2017}]{ricci17}
{Ricci}, C., et~al., 2017, \mnras, 468, 1273

\bibitem[\protect\astroncite{{Risaliti} et~al.}{2009}]{risaliti09}
{Risaliti}, G., et~al., 2009, \mnras, 393, L1

\bibitem[\protect\astroncite{{Rivers} et~al.}{2015a}]{rivers15b}
{Rivers}, E., et~al., 2015a, \apj, 815, 55

\bibitem[\protect\astroncite{{Rivers} et~al.}{2015b}]{rivers15}
{Rivers}, E., et~al., 2015b, \apj, 804, 107

\bibitem[\protect\astroncite{{Sanders} et~al.}{1988}]{sanders88}
{Sanders}, D.~B., {Soifer}, B.~T., {Elias}, J.~H., {Madore}, B.~F., {Matthews},
  K., {Neugebauer}, G., \& {Scoville}, N.~Z.,  1988, \apj, 325, 74

\bibitem[\protect\astroncite{{Satyapal} et~al.}{2014}]{satyapal14}
{Satyapal}, S., {Ellison}, S.~L., {McAlpine}, W., {Hickox}, R.~C., {Patton},
  D.~R., \& {Mendel}, J.~T.,  2014, \mnras, 441, 1297

\bibitem[\protect\astroncite{{Sekiguchi} \& {Wolstencroft}}{1992}]{sekiguchi92}
{Sekiguchi}, K., \& {Wolstencroft}, R.~D.,  1992, \mnras, 255, 581

\bibitem[\protect\astroncite{{Shemmer} et~al.}{2006}]{shemmer06}
{Shemmer}, O., {Brandt}, W.~N., {Netzer}, H., {Maiolino}, R., \& {Kaspi}, S.,
  2006, \apjl, 646, L29

\bibitem[\protect\astroncite{{Silverman} et~al.}{2011}]{silverman11}
{Silverman}, J.~D., et~al., 2011, \apj, 743, 2

\bibitem[\protect\astroncite{{Stern} et~al.}{2012}]{stern12}
{Stern}, D., et~al., 2012, \apj, 753, 30

\bibitem[\protect\astroncite{{Str{\"u}der} et~al.}{2001}]{struder01}
{Str{\"u}der}, L., et~al., 2001, \aap, 365, L18

\bibitem[\protect\astroncite{{van Dam} et~al.}{2006}]{vandam06}
{van Dam}, M.~A., et~al., 2006, \pasp, 118, 310

\bibitem[\protect\astroncite{{Van Wassenhove} et~al.}{2012}]{vanwassenhove12}
{Van Wassenhove}, S., {Volonteri}, M., {Mayer}, L., {Dotti}, M., {Bellovary},
  J., \& {Callegari}, S.,  2012, \apjl, 748, L7

\bibitem[\protect\astroncite{{Vazdekis} et~al.}{2012}]{vazdekis12}
{Vazdekis}, A., {Ricciardelli}, E., {Cenarro}, A.~J., {Rivero-Gonz{\'a}lez},
  J.~G., {D{\'{\i}}az-Garc{\'{\i}}a}, L.~A., \& {Falc{\'o}n-Barroso}, J.,
  2012, \mnras, 424, 157

\bibitem[\protect\astroncite{{Weisskopf}}{1999}]{weisskopf99}
{Weisskopf}, M.~C.,  1999, ArXiv Astrophysics e-prints

\bibitem[\protect\astroncite{{Wizinowich} et~al.}{2006}]{wizinowich06}
{Wizinowich}, P.~L., et~al., 2006, \pasp, 118, 297

\bibitem[\protect\astroncite{{Woods} \& {Geller}}{2007}]{woods07}
{Woods}, D.~F., \& {Geller}, M.~J.,  2007, \aj, 134, 527

\bibitem[\protect\astroncite{{Wright} et~al.}{2010}]{wright10}
{Wright}, E.~L., et~al., 2010, \aj, 140, 1868

\end{thebibliography}

\end{document}